\documentclass[preprint,amsmath,amssymb,aps,pra]{revtex4-1}
\usepackage[utf8]{inputenc}
\usepackage{amsmath}
\usepackage{graphicx}
\usepackage{esint}

\makeatletter
\usepackage{float}
\usepackage{dcolumn}
\usepackage{caption}
\usepackage{bm}
\usepackage{color}
\usepackage{upgreek}

\newcommand{\blue}[1]{{ %\color{blue}
 #1}}
\newcolumntype{L}[1]{>{\raggedright\let\newline\\\arraybackslash\hspace{0pt}}m{#1}}
\newcolumntype{C}[1]{>{\centering\let\newline\\\arraybackslash\hspace{0pt}}m{#1}}
\newcolumntype{R}[1]{>{\raggedleft\let\newline\\\arraybackslash\hspace{0pt}}m{#1}}

\newcommand{\threej}[6]{
\begingroup
\footnotesize{
\renewcommand*{\arraystretch}{0.6}
\begin{pmatrix}
   #1 & #2 & #3 \\
   #4 & #5 & #6 
\end{pmatrix}}
\endgroup
}

\makeatother

\begin{document}

\title{Nonadiabatic phenomena in molecular vibrational polaritons}

\author{Tamás Szidarovszky}
\email{tamas821@caesar.elte.hu}

\affiliation{Institute of Chemistry, ELTE Eötvös Loránd University and MTA-ELTE
Complex Chemical Systems Research Group, H-1117 Budapest, Pázmány
Péter sétány 1/A, Hungary}

\author{Péter Badankó}

\affiliation{Department of Theoretical Physics, University of Debrecen, P.O. Box
400, H-4002 Debrecen, Hungary}

\author{Gábor J. Halász}

\affiliation{Department of Information Technology, University of Debrecen, P.O. Box
400, H-4002 Debrecen, Hungary}

\author{Ágnes Vibók}
\email{vibok@phys.unideb.hu}

\affiliation{Department of Theoretical Physics, University of Debrecen, P.O. Box
400, H-4002 Debrecen, Hungary and ELI-ALPS, ELI-HU Non-Profit Ltd.,
Dugonics tér 13, H-6720 Szeged, Hungary}

\date{\today}
\begin{abstract}
Nonadiabatic phenomena are investigated in the rovibrational motion
of molecules confined in an infrared cavity. Conical intersections
(CIs) between vibrational polaritons, similar to CIs between electronic
polaritonic surfaces, are found. The spectral, topological, and dynamic
properties of the vibrational polaritons show clear fingerprints of
nonadiabatic couplings between molecular vibration, rotation and the
cavity photonic mode. Furthermore, it is found that for the investigated
system, composed of two rovibrating HCl molecules and the cavity mode,
breaking the molecular permutational symmetry, by changing $^{35}$Cl
to $^{37}$Cl in one of the HCl molecules, the polaritonic surfaces,
nonadiabatic couplings, and related spectral, topological, and dynamic
properties can deviate substantially. This implies that the natural
occurrence of different molecular isotopologues needs to be considered
when modeling realistic polaritonic systems. 
\end{abstract}
\maketitle
%abstract: egy szinttel lejjebb, rezgesi poolaritonoknal keressuk a nemadiabatikus effektusok fingerprintjeit. Topologia, spektrum, dinamika. Bemutatjuk HCl-en, ami remek pelda erre.

%\tableofcontents

\section{\label{Introduction}Introduction}

Cavity quantum electrodynamics (CQED) deals with the interaction of
matter with a quantized light field enclosed in a cavity \cite{Miller_2005,Walls_2008}.
Experimental and theoretical studies have revealed that the quantum
nature of the light can significantly modify the different chemical,
physical or energetic properties of material structures \cite{Ruggenthaler_2018,Herrera_2020,cavity_exp_Hutchison_AngChem_2012,cavity_Ebbesen_AccChemRes_2016,Hertzog2019,cavity_Galego_PRX_2015,cavity_Galego_NatCommun_2016,cavity_Feist_ACSphotonics_2018,Schafer_2018,cavity_Herrera_PRL_2016,dark_vibronic_polaritons_Herrera_PRL_2017,Markus_1a,cavity_Kowalewski_JPCL_2016,Davidsson_2020,Mandal_2019,Joel_1a,cavity_Luk_JCTC_2017,Gerit_2a,Tamas_1a,Agi_2,Csehi_2019b,cavity_MCTDH_Vendrell_CP_2018,Oriol_2a,Gu_2020,Pino_2015,Dunkelberger_2016,theory_of_organic_cavities_Herrera_ACSphotonics_2018,Ahn_2018,Ebbesen_3a,Ebbesen_4a,Ebbesen_5a,Ebbesen_7a,Thomas_2019,Long_2014,cavity_Muallem_JPCL_2016,Fleischer_2019,Joel_4a,Joel_3a,Mandal2020,Ulusoy2019,Du2019,Gu2020_JPCL,Szidarovszky2020}.
Most of the works consider only a single confined photonic mode in
an optical or plasmonic cavity which can strongly interact with atoms,
molecules or nano materials. The strong coupling regime is reached
when the light–matter coupling is stronger than the loss of the cavity
mode and the decay rates in the matter. In the case of molecules,
the confined photonic mode of the cavity can efficiently couple both
electronic or vibrational states, depending on the cavity mode wavelength.
In the first situation the different electronic molecular states mix
with the UV–visible cavity photon forming hybrid light–matter, so-called
electronic or vibronic polariton states carrying both characters of matter and light.
Several experimental %\cite{cavity_exp_Hutchison_AngChem_2012,cavity_Ebbesen_AccChemRes_2016}
and theoretical %\cite{Ruggenthaler_2018,Herrera_2020,cavity_Galego_PRX_2015,cavity_Galego_NatCommun_2016,cavity_Feist_ACSphotonics_2018,Schafer_2018,cavity_Herrera_PRL_2016,dark_vibronic_polaritons_Herrera_PRL_2017,Markus_1a,cavity_Kowalewski_JPCL_2016,Davidsson_2020,Mandal_2019,Joel_1a,cavity_Luk_JCTC_2017,Gerit_2a,Tamas_1a,Agi_2,Csehi_2019b,cavity_MCTDH_Vendrell_CP_2018,Oriol_2a,Gu_2020}
studies have demonstrated that this phenomenon gives rise to a variety
of interesting effects such as modification of the chemical landscapes
\cite{cavity_exp_Hutchison_AngChem_2012,cavity_Ebbesen_AccChemRes_2016,cavity_Galego_NatCommun_2016,Joel_1a},
cavity-induced nonadiabatic phenomena \cite{Ruggenthaler_2018,cavity_Galego_PRX_2015,cavity_Feist_ACSphotonics_2018,Markus_1a,cavity_Kowalewski_JPCL_2016,Davidsson_2020,Tamas_1a,Agi_2,Csehi_2019b,cavity_MCTDH_Vendrell_CP_2018,Oriol_2a,Gu_2020},
controlling photochemistry \cite{Markus_1a,cavity_Kowalewski_JPCL_2016,Mandal_2019,cavity_Luk_JCTC_2017,Gerit_2a,Gu_2020}\blue{, or intermolecular collective effects \cite{Mandal_2019,Ulusoy2019,Du2019,Gu2020_JPCL,Szidarovszky2020}}.

Recently efforts have been made to study the vibrational strong coupling
(VSC) within one molecular electronic state \cite{Pino_2015,theory_of_organic_cavities_Herrera_ACSphotonics_2018,Long_2014,cavity_Muallem_JPCL_2016,Ribeiro2017,Joel_4a,Joel_3a}.
In this regime experiments focus on the cavity-induced manipulation
of chemical reactions in the ground electronic state or studying the
effects of vibrational polaritons in solid and liquid phase inside
infrared (IR) Fabry–Pérot cavities \cite{Dunkelberger_2016,Ahn_2018,Ebbesen_3a,Ebbesen_4a,Ebbesen_5a,Ebbesen_7a,Thomas_2019}.
The presence of the vibrational strong coupling may lead to the suppression
or enhancement of reactive pathways for molecules even in the absence
of external photon pumping, implying that they involve thermally activated
processes \cite{Ebbesen_7a,Thomas_2019,Joel_4a,Joel_3a}. \blue{As a necessary initial step to understand chemical reactivity, the detailed theoretical description and characterization of individual molecular vibrations interacting with an IR cavity mode also received focus \cite{Hernandez2019,Triana2020}}. 

In the present work we extend the theoretical approach by incorporating
the effect of molecular rotation with vibrational polaritons. We employ
realistic and highly-accurate molecular models, based on accurate
quantum chemical and variational nuclear motion computations. This
allows for a high-resolution insight into the formation, the spectroscopy
and the dynamics of rovibrational polaritons, including cavity photon-mediated
energy transfer between the different types of molecules. Because
utilizing microfluidic cavities allows for the experimental realization
of vibrational polaritons in the liquid phase, and because gas phase
molecular polaritons are likely to be investigated eventually, special
emphasis is given to the role of molecular rotations. The coupling
strength between the cavity mode and the vibrational transition of
the molecule naturally depends on the spatial orientation of the molecule.
If molecular rotation is feasible, such as in the liquid or gas phase,
then a simple averaging over different orientations is in principle
erroneous, because the orientation dependent coupling strength couples
the rotational degrees of freedom with the vibrational ones. Such
couplings give rise to nonadiabatic effects \cite{Cederbaum_multimode,Yarkony_1996,Baer_2002,vibronic_coupling_model_Cederbaum_AnnRevPhysChem_2004,Domcke_2004,Baer_2006}
or in some cases to the formation of light-induced conical intersections
(LICIs) \cite{Nimrod_1,LICI1,LICI2} , which can have a significant
impact on the polaritonic properties \cite{Tamas_1a,Agi_2}.
\blue{The impact of simple orientational averaging has been investigated in the context of light-induced electronic conical intersections for example in Refs. \cite{LICI3,LICI5,LICI_in_spectrum_Szidarovszky_JPCL_2018}. It is difficult to judge in general the error introduced by neglecting nonadiabatic coupling between rotations and vibrations in systems with vibrational strong coupling, because the error is strongly system specific. Nonetheless, it is an issue which should be considered.}

We introduce the concept of vibrational potential energy surfaces,
on which rotational dynamics proceed, and show the impact of the cavity-induced
nonadiabatic coupling – between rotations and vibrations – on spectroscopic,
dynamical and topological properties of molecules. Our showcase system
is a mixture of H$^{35}$Cl and H$^{37}$Cl molecules interacting
with a photonic mode.

%The article is organized as follows. In the next section, we present the theoretical background and computational details required for our study. In the third section the results and discussions are reported. In the last section, we summarize the conclusions.

\section{\label{Theory}Theoretical approach}

\subsection{Rovibrational polaritons}

The hamiltonian of rotating-vibrating molecules interacting with a
single lossless cavity mode can be written in the dipole approximation
as 
\begin{equation}
\begin{split}\hat{H}=\sum_{i=1}^{N_{{\rm mol}}}(\hat{H}_{{\rm m}}^{(i)}-\hat{{\rm \mathbf{E}}}_{{\rm c}}\hat{\mathbf{\upmu}}^{(i)})+\hat{H}_{{\rm c}}=\\
\sum_{i=1}^{N_{{\rm mol}}}\big(\hat{H}_{{\rm m}}^{(i)}-\sqrt{\frac{\hbar\omega_{{\rm c}}}{\varepsilon_{0}V}}\mathbf{e}\hat{\mathbf{\upmu}}^{(i)}(\hat{a}_{{\rm c}}^{\dagger}+\hat{a}_{{\rm c}})\big)+\hbar\omega_{{\rm c}}\hat{a}_{{\rm c}}^{\dagger}\hat{a}_{{\rm c}}.
\end{split}
\label{eq:hamiltonian}
\end{equation}
where $\hat{H}_{{\rm m}}^{(i)}$ is the \textit{i}th field-free molecular
rovibrational Hamiltonian, $\hat{a}_{c}^{\dagger}$ and $\hat{a}_{c}$
are photon creation and annihilation operators, respectively, $\omega_{c}$
is the frequency of the cavity mode, $\hslash$ is Planck's constant
divided by $2\pi$, $\varepsilon_{0}$ is the electric constant, $V$
is the volume of the electromagnetic mode, $\mathbf{e}$ is the polarization
vector of the cavity mode, and $\hat{\mathbf{\upmu}}^{(i)}$ is the
dipole moment of the \textit{i}th molecule.

We assume the cavity electric field to be polarized along the lab-fixed
z-axis, $\mathbf{e}=(0,0,1)$, and we consider an IR cavity, which
allows restricting the molecular model to a single electronic state.
Then, the 
\begin{equation}
\vert N\rangle\prod_{i=1}^{N_{{\rm mol}}}\vert\Psi_{{\rm rovib}}^{i,n_{i}J_{i}M_{i}}\rangle\label{eq:direct_product_basis}
\end{equation}
direct-product functions, also called diabatic states, provide a complete
basis, where the $\vert\Psi_{{\rm rovib}}^{i,nJM}\rangle$ field-free
rovibrational eigenstates satisfy 
\begin{equation}
\hat{H}_{{\rm m}}^{(i)}\vert\Psi_{{\rm rovib}}^{i,nJM}\rangle=E_{{\rm rovib}}^{(i),nJ}\vert\Psi_{{\rm rovib}}^{i,nJM}\rangle,\label{eq:field_free_schr_Eq}
\end{equation}
$J$ and $M$ are the rotational angular momentum and its projection
onto the space-fixed z-axis, respectively, $n$ is all other quantum
numbers uniquely defining the rovibrational states, and $\vert N\rangle$
is a photon number state of the cavity radiation. For a diatomic molecule $n$ can be identified with the single vibrational quantum number $v$, and using the notation $\vert\Psi_{{\rm rovib}}^{i,nJM}\rangle\equiv\vert v_{i}J_{i}M_{i}\rangle$, the matrix representation
of Eq. (\ref{eq:hamiltonian}) using the basis of Eq. (\ref{eq:direct_product_basis})
gives matrix elements of the following form: 
\begin{equation}
\begin{split} & \Big(\langle N\vert\prod_{i=1}^{N_{{\rm mol}}}\langle v_{i}J_{i}M_{i}\vert\Big)\hat{H}\Big(\prod_{j=1}^{N_{{\rm mol}}}\vert v'_{j}J'_{j}M'_{j}\rangle\vert N'\rangle\Big)=\\
 & \sum_{i=1}^{N_{{\rm mol}}}\delta_{M_{i}M'_{i}}\Big[E_{{\rm rovib}}^{(i),v_{i}J_{i}}\delta_{v_{i}v'_{i}}\delta_{J_{i}J'_{i}}\delta_{NN'}-\sqrt{\frac{\hbar\omega_{{\rm c}}}{\varepsilon_{0}V}}\langle v_{i}\vert\mu^{(i)}(R_{i})\vert v'_{i}\rangle\langle J_{i}M_{i}\vert{\rm cos}(\theta_{i})\vert J'_{i}M_{i}\rangle\times\\
 & \Big(\sqrt{N}\delta_{NN'+1}+\sqrt{N+1}\delta_{NN'-1}\Big)\Big]\prod_{j\neq i}\delta_{v_{j}v'_{j}}\delta_{J_{j}J'_{j}}\delta_{M_{j}M'_{j}}+N\hbar\omega_{{\rm c}}\delta_{NN'}\prod_{j}^{N_{{\rm mol}}}\delta_{v_{j}v'_{j}}\delta_{J_{j}J'_{j}}\delta_{M_{j}M'_{j}},
\end{split}
\label{eq:diatomic_hamiltonian_in_general_basis}
\end{equation}
where $\mu^{(i)}(R_i)$ is the permanent dipole moment as a function
of the internuclear distance $R_{i}$, $\theta_{i}$ is the angle
between the space-fixed z axis and the \textit{i}th molecule, and
the energy scale was set so that the zero point energy of the photonic
mode, $\hbar\omega_{{\rm c}}/2$, is omitted.

The rovibrational polaritonic states are the eigenstates of the Hamiltonian
of Eq. (\ref{eq:hamiltonian}). The \textit{k}th polariton can be
written as 
\begin{equation}
\vert\Psi_{{\rm pol}}^{k}\rangle=\sum_{N}\sum_{v_{i},J_{i},M_{i}}C_{N,v_{1}J_{1}M_{1},...,v_{N_{{\rm mol}}}J_{N_{{\rm mol}}}M_{N_{{\rm mol}}}}^{k}\prod_{i}\vert v_{i}J_{i}M_{i}\rangle\vert N\rangle,\label{eq:polaritonic_wave_function}
\end{equation}
where $C_{N,v_{1}J_{1}M_{1},...,v_{N_{{\rm mol}}}J_{N_{{\rm mol}}}M_{N_{{\rm mol}}}}^{k}$
are the coefficients of the \textit{k}th eigenvector of the matrix
having the elements given in Eq. (\ref{eq:diatomic_hamiltonian_in_general_basis}).

\subsection{Computational details}

The field-free rovibrational energies of the H$^{35}$Cl and H$^{37}$Cl
molecules were computed as $E_{{\rm rovib}}^{(i),v_{i}J_{i}}=E_{{\rm vib}}^{(i),v_{i}}+B_{v_{i}}^{(i)}J(J+1)$,
where the $E_{{\rm vib}}^{(i),v_{i}}$ vibrational energies were obtained
from converged variational computations utilizing the spherical-DVR
discrete variable representation (DVR) \cite{D2FOPI_Szidarovszky_PCCP_2010}
and the potential energy curve (PEC) of Ref. \cite{Na2_PEC_Magnier_JCP_1993}.
The $B_{v}$ rotational constants (given in energy units) were evaluated
for each vibrational state as $B_{v}^{(i)}=\langle v_{i}\vert\hbar^{2}/(2m_{{\rm red}}^{(i)}R^{2})\vert v_{i}\rangle$,
where $m_{{\rm red}}^{(i)}$ is the reduced vibrational mass of the
\textit{i}th molecule. The nuclear masses used in the computations
were $m_{{\rm H}}=1.00784$ \textit{u}, $m_{{\rm ^{35}Cl}}=34.9689$
\textit{u}, and $m_{{\rm ^{37}Cl}}=36.9659$ \textit{u}.

For computing the $\langle v_{i}\vert\mu^{(i)}(R)\vert v'_{i}\rangle \equiv \mu^{(i)}_{v_{i} v'_{i}}$
vibrational (transition) dipole elements, the dipole moment curve
of Ref. \cite{Na2_TDM_Zemke_JMS_1981} was used. \blue{The values obtained are $\mu^{\rm (H^{35}Cl)}_{00}=0.42858$, $\mu^{\rm (H^{35}Cl)}_{10}=\mu^{\rm (H^{35}Cl)}_{01}=0.027055$, $\mu^{\rm (H^{37}Cl)}_{00}=0.42857$, and $\mu^{\rm (H^{37}Cl)}_{10}=\mu^{\rm (H^{37}Cl)}_{01}=0.027045$, all in atomic units.} The $\langle J_{i}M_{i}\vert{\rm cos}(\theta)\vert J'_{i}M_{i}\rangle$
terms were evaluated using an expression involving 3-j symbols,
\begin{equation}
\langle JM\vert{\rm cos}(\theta)\vert J'M'\rangle=\sqrt{(2J+1)(2J'+1)}(-1)^{M'}\threej{J}{1}{J'}{M}{0}{-M'}\threej{J}{1}{J'}{0}{0}{0},
\end{equation}
which can be derived using the relations $\langle\phi,\theta\vert J\,M\rangle=Y_{JM}(\theta,\phi)=\sqrt{\frac{2J+1}{4\pi}}{D_{M0}^{J}}^{*}(\phi,\theta,\chi)$,
${\rm cos}(\theta)={D_{00}^{1}}(\phi,\theta,\chi)$, and $\int D_{K_{1}M_{1}}^{J_{1}}(\mathbf{\Omega})D_{K_{2}M_{2}}^{J_{2}}(\mathbf{\Omega})D_{K_{3}M_{3}}^{J_{3}}(\mathbf{\Omega})d\mathbf{\Omega}=8\pi^{2}\threej{J_{1}}{J_{2}}{J_{3}}{K_{1}}{K_{2}}{K_{3}}\threej{J_{1}}{J_{2}}{J_{3}}{M_{1}}{M_{2}}{M_{3}}$
\cite{88Zare}, where $D_{KM}^{J}(\mathbf{\Omega})$ is the Wigner
D-matrix and $\mathbf{\Omega}$ represents the three Euler angles
$\phi$, $\theta$, and $\chi$.

%   \subsubsection{Polaritonic states and quantum dynamics}

\section{Results and discussion}

\subsection{Vibrational polaritonic energy surfaces (VPES)}

For understanding the underlying physics, it is useful to construct
the matrix representation of the Hamiltonian in Eq. (\ref{eq:hamiltonian})
using vibrational-photonic basis functions $\vert N\rangle\prod_{i=1}^{N_{{\rm mol}}}\vert\Psi_{{\rm vib}}^{i,v_{i}}\rangle$.
The matrix elements of the resulting vibro-photonic Hamiltonian depend
on the $\theta$ rotational coordinates. After removing the rotational
kinetic energy terms, the eigenvalues of the vibro-photonic Hamiltonian
form vibrational polaritonic energy surfaces (VPES), which are functions
of the $N_{{\rm mol}}$ rotational coordinates. Note the analogy that
the VPESs provide effective potential energy surfaces (PES) for the
rotational motion the same way as the usual light-dressed electronic
potentials provide effective PESs for vibrational motion in polyatomics.

For two diatomic molecules (denoted as molecule 1 and 2, respectively)
coupled to a single IR cavity mode, the first 4$\times$4 block in
the vibro-photonic Hamiltonian reads 
\begin{equation}
\mathbf{H}^{4\times4}=\mathbf{T}_{{\rm rot}}+\mathbf{V}_{{\rm rot}},
\end{equation}
where \begin{footnotesize} 
\begin{equation}
\mathbf{T}_{{\rm rot}}=\begin{bmatrix}B_{0}^{(1)}\hat{T}_{{\rm rot}}^{(1)}+B_{0}^{(2)}\hat{T}_{{\rm rot}}^{(2)} & 0 & 0 & 0\\
0 & B_{0}^{(1)}\hat{T}_{{\rm rot}}^{(1)}+B_{0}^{(2)}\hat{T}_{{\rm rot}}^{(2)} & 0 & 0\\
0 & 0 & B_{1}^{(1)}\hat{T}_{{\rm rot}}^{(1)}+B_{0}^{(2)}\hat{T}_{{\rm rot}}^{(2)} & 0\\
0 & 0 & 0 & B_{0}^{(1)}\hat{T}_{{\rm rot}}^{(1)}+B_{1}^{(2)}\hat{T}_{{\rm rot}}^{(2)}
\end{bmatrix},
\end{equation}

\begin{equation}
\begin{split}\mathbf{V}_{{\rm rot}} & =\begin{bmatrix}E_{{\rm vib}}^{(1),0}+E_{{\rm vib}}^{(2),0} & 0 & 0 & 0\\
0 & E_{{\rm vib}}^{(1),0}+E_{{\rm vib}}^{(2),0}+\hbar\omega_{{\rm c}} & 0 & 0\\
0 & 0 & E_{{\rm vib}}^{(1),1}+E_{{\rm vib}}^{(2),0} & 0\\
0 & 0 & 0 & E_{{\rm vib}}^{(1),0}+E_{{\rm vib}}^{(2),1}
\end{bmatrix}-\sqrt{\hbar\omega_{{\rm c}}/(\varepsilon_{0}V)}\times\\
 & \begin{bmatrix}0 & \mu_{00}^{(1)}{\rm cos}(\theta_{1})+\mu_{00}^{(2)}{\rm cos}(\theta_{2}) & 0 & 0\\
\mu_{00}^{(1)}{\rm cos}(\theta_{1})+\mu_{00}^{(2)}{\rm cos}(\theta_{2}) & 0 & \mu_{01}^{(1)}{\rm cos}(\theta_{1}) & \mu_{01}^{(2)}{\rm cos}(\theta_{2})\\
0 & \mu_{10}^{(1)}{\rm cos}(\theta_{1}) & 0 & 0\\
0 & \mu_{10}^{(2)}{\rm cos}(\theta_{2}) & 0 & 0
\end{bmatrix},
\end{split}
\label{eq:VPES_hamiltonian}
\end{equation}
\end{footnotesize} where $\mu_{kl}^{(i)}=\langle v_{i}=k\vert\mu^{(i)}(R)\vert v_{i}=l\rangle$,
$\hat{T}_{{\rm rot}}^{(i)}$ is the squared rotational angular momentum of the \textit{i}th molecule, and $B_{v}^{(i)}$ is the rotational constant of the \textit{i}th molecule in its \textit{v}th vibrational state. For the specific case of a H$^{35}$Cl
and a H$^{37}$Cl molecule, Figure \ref{fig:VPESs} show the
three largest eigenvalues of the $\mathbf{V}_{{\rm rot}}$ matrix
in Eq. (\ref{eq:VPES_hamiltonian}) as a function of the $\theta_{i}$
variables for a coupling strength of $ \mu^{\rm (H^{35}Cl)}_{00} \sqrt{\hbar\omega_{{\rm c}}/(\varepsilon_{0}V)}=33.26$ cm$^{-1}$. 
The same is shown in Fig. \ref{fig:VPESs_2} for the case of two identical H$^{35}$Cl molecules. 
\blue{
Fig. \ref{fig:VPESs} demonstrates that in the case of the two different
isotopologues, three polaritons are formed, as expected \cite{Houdr1996,Szidarovszky2020},
and conical intersections (CI) can be formed between the VPESs. In
the case of two identical H$^{35}$Cl molecules, Fig. \ref{fig:VPESs_2}
shows that the middle polaritonic surface is flat, which one would usually identify as a dark polaritonic state (although it is not dark in this case, see below). The upper (lower) VPES touches the dark state for red (blue) shifted cavity
wave lengths.
}

\subsection{Topological properties}

A well-known consequence of CIs or LICIs between adiabatic electronic
PESs is that the electronic wave functions of the connected surfaces
become double valued, i.e., propagating the electronic wave function
of a single surface along a closed curve around the CI leads to the
accumulation of a topological phase of $\pi$ \cite{Longuet-Higgins1958,Longuet-Higgins1961429,Mead1982,Berry1984,LICI1,Halsz2018}.
In order to verify that the LICIs between VPESs are indeed true LICIs
(and to confirm the existence of LICIs between VPESs even when the
diagonal dipole terms are not omitted from the VPESs), the numerical
verification of the topological phases was carried out. Now, we borrow
a well-established %nomenclature
framework developed for natural nonadiabatic phenomena,
which was succesfully applied to study the topological properties
of ``natural'' electronic degeneracies \cite{Baer1975112,Baer_2002,Vertesi20053476,Baer_2006}.

Let us start from the adiabatic-to-diabatic transformation (ADT)
\begin{equation}
\mathbf{W}=\mathbf{A}^{\dagger}\mathbf{VA}\label{eq:ADT}
\end{equation}
where $\mathbf W$ and $\mathbf V$ denote the diabatic and adiabatic PES, respectively.
$\mathbf A$ is a unitary matrix called ADT matrix, which can be obtained by solving the equation
\begin{equation}
\nabla \mathbf A+ \bm{\tau} \mathbf A=0.\label{eq:Baer-eq}
\end{equation}
Its solution can be written in the following form 
\begin{equation}
\mathbf A(s)=\gamma\,exp\left(-\int_{s_{0}}^{s}\bm{\tau}(s)\,ds\right)\mathbf A(s_{0})\label{eq:A-mat}
\end{equation}
where $\gamma$ is a path ordering operator and $\bm{\tau}(s)$ is the
nonadiabatic coupling matrix which contains the nonadibatic coupling
terms (NAC)s
\begin{equation}
\bm{\tau}_{i,j}=<\Psi_{i}|\nabla|\Psi_{j}>.\label{eq:tau}
\end{equation}
Here $\Psi_{i}$ and $\Psi_{j}$ are $ith$ and $jth$ adiabatic vibrational
wave functions, respectively and $\nabla$ is the gradient operator
corresponding to the rotational coordinates.
%\red{EZ NEKEM MOST NEM VILAGOS TELJESEN. KLASSZIKUS ESETBEN VAN REZGESI KOORDINATAK SZERINTI GRADIENS ES ELEKTRONALLAPOT HA JOL EMLEKSZEM. ITT PEDIG REZGESI ALLAPOT ES FORGASI KOORDINATA TERBELI GRADIENS HA JOL GONDOLOM. EZT TALAN LEHETNE KICSIT KONKRETABBAN LEIRNI}.

The necessary condition for the $\mathbf A$-matrix to yield single-valued
diabatic wave function is that the $\mathbf D$-matrix,
\begin{equation}
D(\Gamma)=\gamma\,exp\left(-\oint_{\Gamma}\bm{\tau}\,ds\right)\label{eq:D-matrix}
\end{equation}
needs to be diagonal and has, in its diagonal, numbers of ($+1$)s
or ($-1$)s. The number of ($-1$)s in the $\mathbf D$ is designated as $K$
which is the number of the adiabatic vibrational eigenfunctions that
flip their sign. The $\mathbf D$-matrix is the topological matrix which contains
all topological features of a vibrational manifold in a region surrounded
by its contour $\varGamma$ and $K$ is the topological number. The
positions of the ($-1$)s in the $\mathbf D$-matrix correspond with the vibrational
eigenfunctions that flip their sign. The equation for $\mathbf D(\Gamma)$
is a kind of quantization condition to be fulfilled by the nonadiabatic
coupling matrix $\bm{\tau}$ along any closed contour $\varGamma$ in configuration
space (CP).

\begin{figure}[h!]
\centering \includegraphics[width=0.45\textwidth]{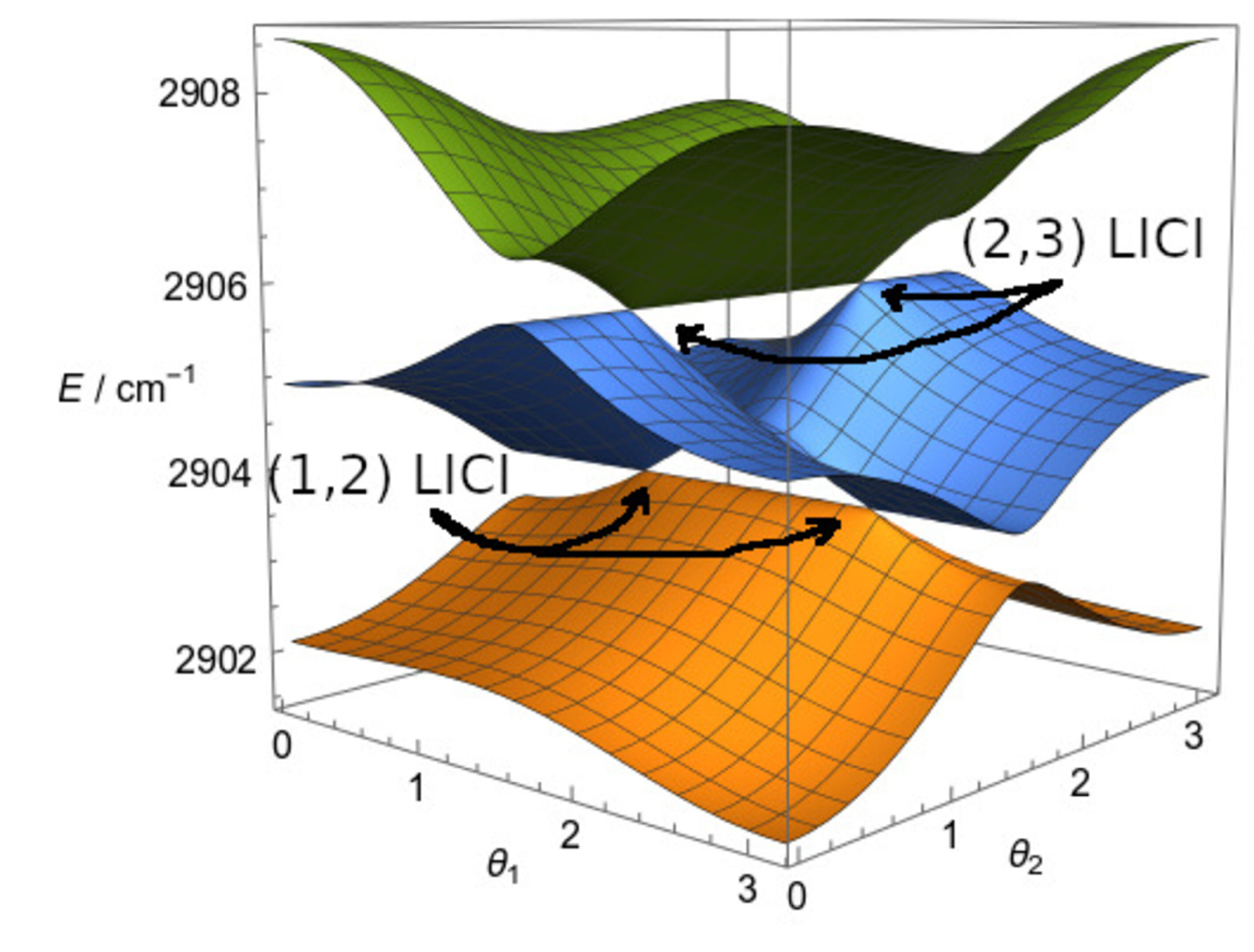}

\caption{Vibrational polaritonic energy surfaces (VPES) of the ``H$^{35}$Cl
+ H$^{37}$Cl + IR cavity mode'' system, obtained by diagonalizing
the $\mathbf{V}_{{\rm rot}}$ matrix in Eq. (\ref{eq:VPES_hamiltonian})
using the parameters
$E_{{\rm vib}}^{1,0}+E_{{\rm vib}}^{2,0}+\hbar\omega_{{\rm c}}=2904.5$
cm$^{-1}$, $E_{{\rm vib}}^{1,1}+E_{{\rm vib}}^{2,0}=2905.9$ cm$^{-1}$,
and $E_{{\rm vib}}^{1,0}+E_{{\rm vib}}^{2,1}=2903.7$ cm$^{-1}$.
The $\mu^{\rm (H^{35}Cl)}_{00}\sqrt{\hbar\omega_{{\rm c}}/(\varepsilon_{0}V)}$ coupling strength
is 33.26 cm$^{-1}$. Two types of light-induced conical intersections (LICI) are also highlighted, see text.}
\label{fig:VPESs} 
\end{figure}

\begin{figure}[h!]
\centering \includegraphics[width=0.45\textwidth]{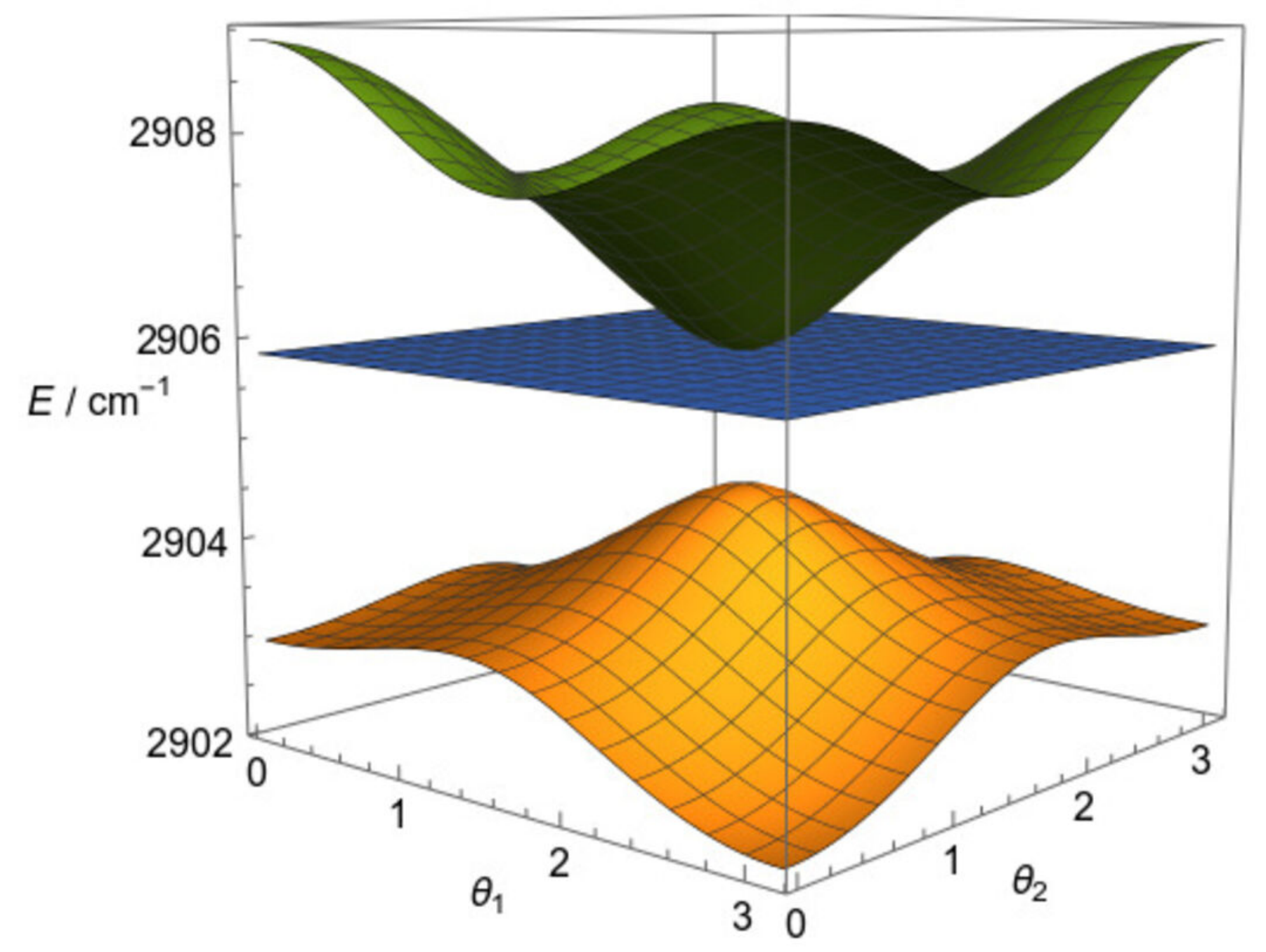}

\caption{Same as Fig. \ref{fig:VPESs}, but for two identical H$^{35}$Cl molecules
in the cavity. The parameters are $E_{{\rm vib}}^{1,1}+E_{{\rm vib}}^{2,0}=E_{{\rm vib}}^{1,0}+E_{{\rm vib}}^{2,1}=2905.9$
cm$^{-1}$, $\mu^{\rm (H^{35}Cl)}_{00}\sqrt{\hbar\omega_{{\rm c}}/(\varepsilon_{0}V)}=33.26$
cm$^{-1}$, while $E_{{\rm vib}}^{1,0}+E_{{\rm vib}}^{2,0}+\hbar\omega_{{\rm c}}=2904.5$
cm$^{-1}$. }
\label{fig:VPESs_2} 
\end{figure}

In case of two quasi-isolated adiabatic states ($j,j+1$) the quantization
condition reads:
\begin{equation}
\alpha_{j,j+1}(\Gamma)=\oint_{\Gamma}\bm{\tau}_{j,j+1}(s|\Gamma)\,ds=n_{j}\,\pi\label{eq:alpha1}
\end{equation}
where $\alpha_{j,j+1}(\Gamma)$ is the topological phase. 
%\red{MIT JELENT A $(s|q)$?}
When the closed
contour $\Gamma$ is a circle with radii q:
\begin{equation}
\alpha_{j,j+1}(q)=\int_{0}^{2\pi}\bm{\tau}_{j,j+1}(\varphi|q)d\varphi.\label{eq:alpha2}
\end{equation}
The adiabatic-to-diabatic transformation angle can be defined as:
\begin{equation}
\gamma_{j,j+1}(\varphi|q)=\int_{0}^{\varphi}\bm{\tau}_{\varphi j,j+1}(\varphi|q)d\varphi.\label{eq:gamma}
\end{equation}
In the related numerical calculations of this study we concentrate on LICIs between the lower and
middle as well as the middle and upper vibrational polaritonic states
(see Figs. \ref{fig:VPESs} and \ref{fig:VPESs_2}). In Figure \ref{fig:VPESs}
these are the two (1,2) and two (2,3) LICIs, corresponding to the
three bright states, which are formed for the ``H$^{35}$Cl + H$^{37}$Cl
+ IR cavity mode'' system. These are ``true'' LICIs, i.e., first-order
degeneracy points, as the two HCl molecules are distinguishable. However,
for the two indistinguishable confined H$^{35}$Cl molecules, only
a second-order degeneracy point is formed, as demonstrated in Fig.
\ref{fig:VPESs_2}. 

\begin{figure}[h!]
\centering \includegraphics[width=0.45\textwidth]{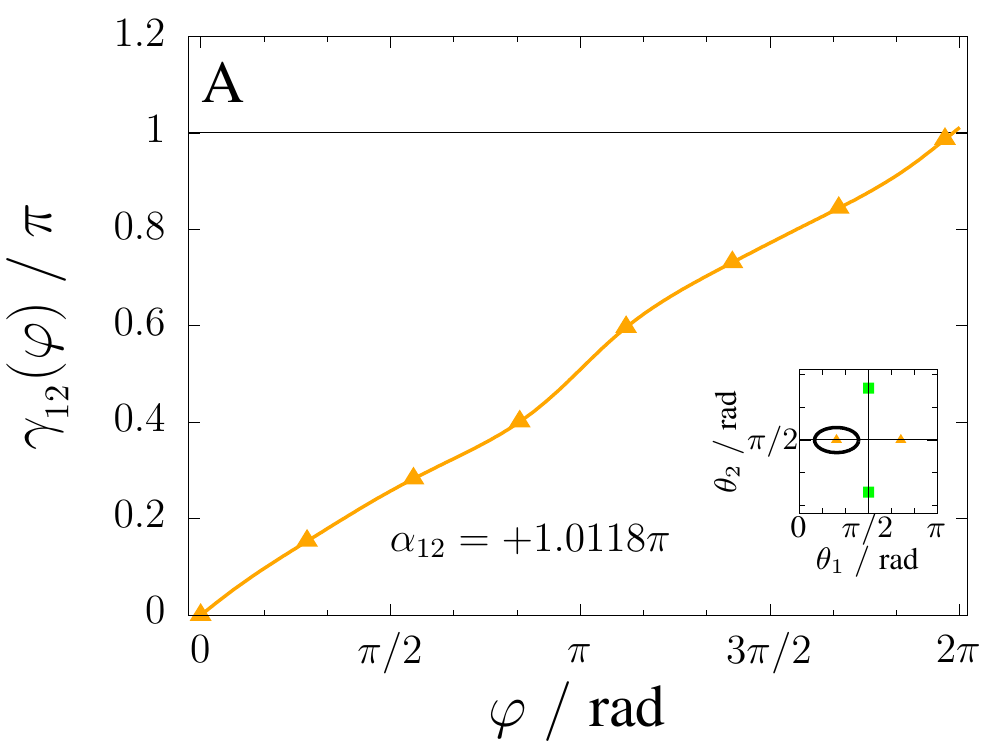} \includegraphics[width=0.45\textwidth]{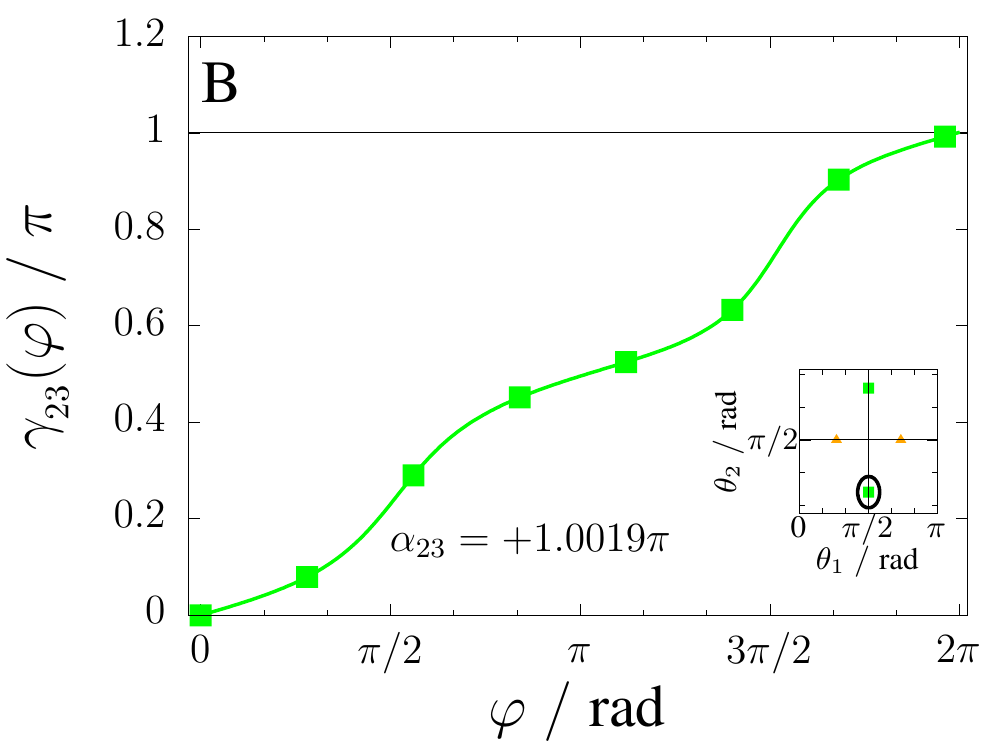}
\includegraphics[width=0.45\textwidth]{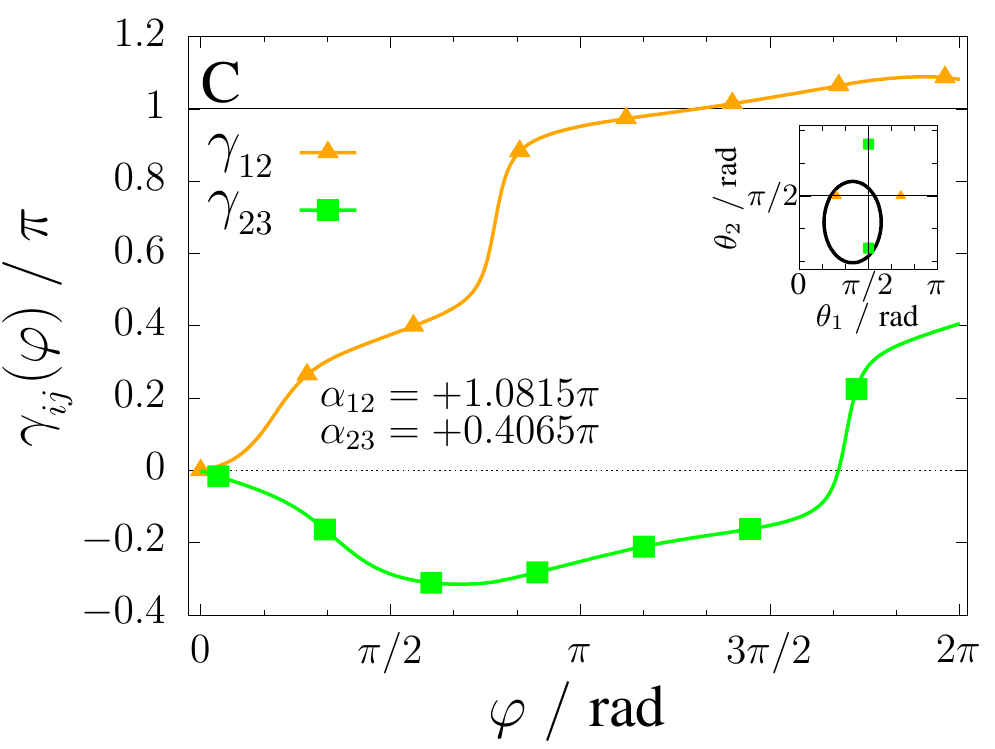} \includegraphics[width=0.45\textwidth]{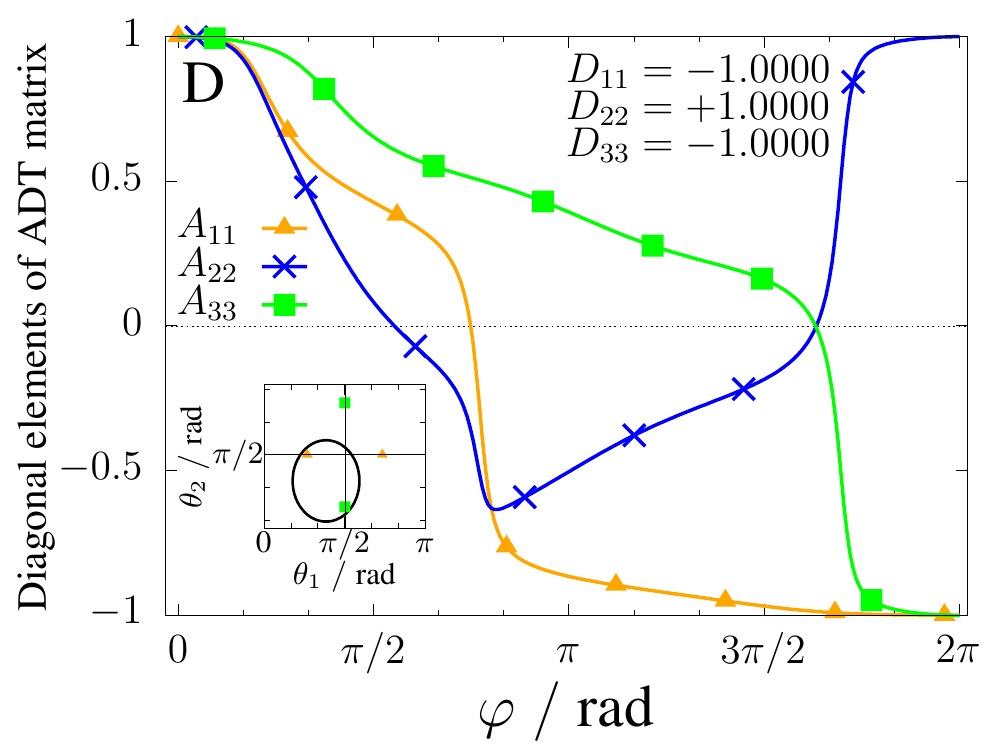}

\caption{The adiabatic-to-diabatic transformation angle $\gamma_{i,j}$ (on panels A, B and C) and
the diagonal elements of the A-topological matrix (on panel D) as a function of
$\varphi\in[0,2\pi]$. $\varphi$ defines the position on the closed
curve. The topological phase $\alpha_{i,j}$ and the $D$-topological
matrix elements are both shown for the ``H$^{35}$Cl + H$^{37}$Cl
+ IR cavity mode'' system. }
. \label{fig:Berry_1} 
\end{figure}

In Fig. \ref{fig:Berry_1} we report on results for the ``H$^{35}$Cl
+ H$^{37}$Cl + IR cavity mode'' system. At first we calculated the
$\bm{\tau}_{1,2}$ along a circle located at the (1,2)LICI point. In Fig.
\ref{fig:Berry_1}. panel A the related ADT angle $\gamma_{1,2}$
and the corresponding topological phase $\alpha_{1,2}=-1.0169\pi$
are shown. Similar quantities are presented for the (2,3)LICI in panel
B. Here we obtain $\alpha_{2,3}=1.0077\pi$. It is well noticed that
both $\alpha_{1,2}$ and $\alpha_{2,3}$ are close to $\pi$ as expected.
This means that there is only
one (1,2)LICI and one (2,3)LICI in the studied CS region of panel A and B, respectively, and the
two-state approximation works well. In panel C the closed contour,
in contrast to the two previous ones, surrounds two LICI points.
The calculated $\alpha_{1,2}=-1.1227\pi$ and $\alpha_{2,3}=-0.3436\pi$
topological phases are far from being close to $\pi$, indicating that
the two-state approximation fails to describe accurately the topological
behavior. The third state is needed to be involved in the calculations
for the proper topological description. In order to derive the three-state
ADT matrix elements and the corresponding $\mathbf D$ matrix one has to calculate
all off-diagonal matrix elements of the $\bm{\tau}(s)$ NAC matrix and
then apply Eqs. \ref{eq:A-mat} and \ref{eq:D-matrix}. On panel D of Fig. \ref{fig:Berry_1} the values for the $\mathbf D$ matrix elements are presented for the closed curve indicated as an inset in the panel.
%\red{A KOVETKEZO MONDAT NEKEM ZAVAROS} 
The values of the diagonal elements ($-1$,$+1$,$-1$) clearly imply that the 
distribution contains an odd number (presumably one) of LICI for each type (1,2)LICI and (2,3)LICI.
Therefore, the first and third vibrational wave functions change their
sign while the sign of the second vibrational wave function remains
unchanged after flipping its sign twice. 

\begin{figure}[h!]
\centering \includegraphics[width=0.45\textwidth]{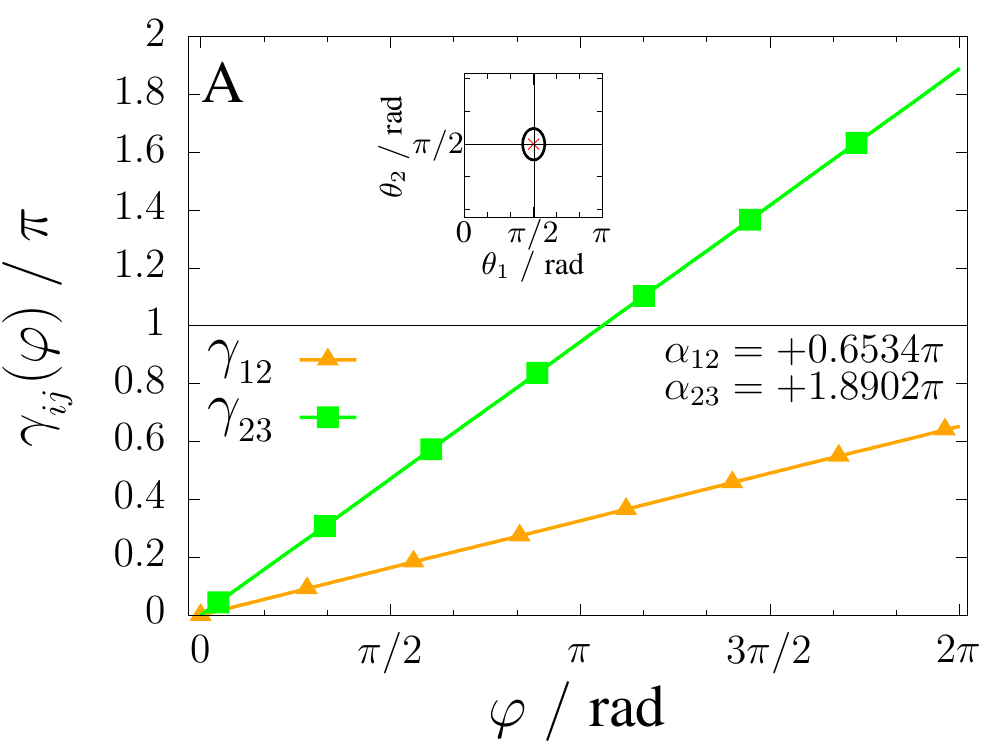}
\includegraphics[width=0.45\textwidth]{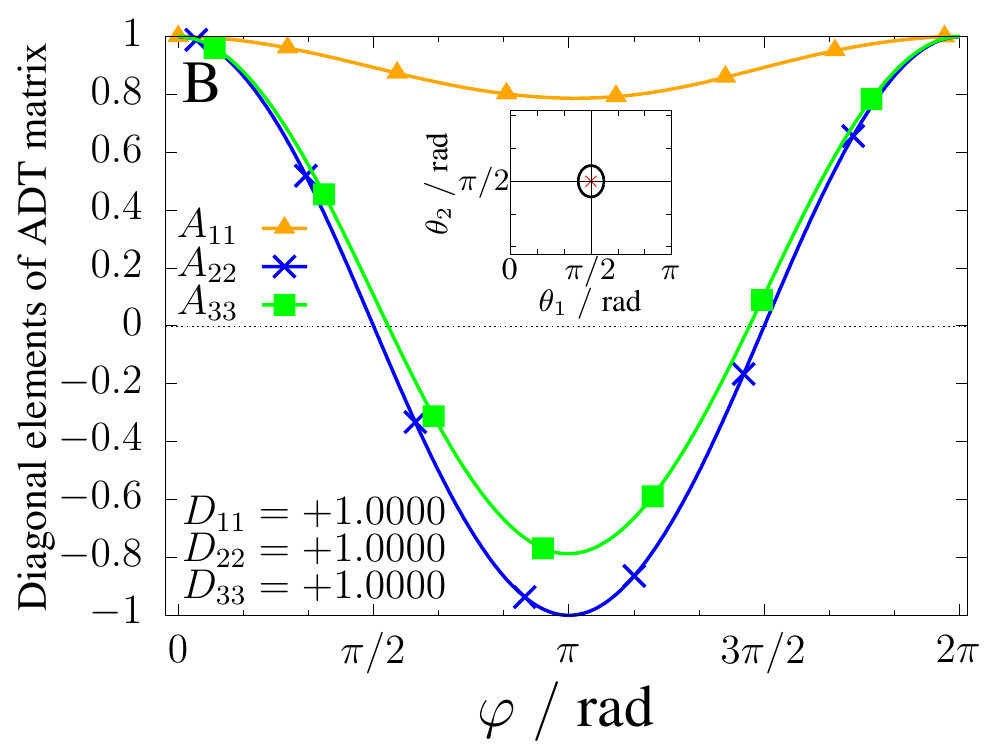} \caption{The adiabatic-to-diabatic transformation angle $\gamma_{i,j}$ (panel A) and
the diagonal elements of the A-topological matrix (panel B) as a function of
$\varphi\in[0,2\pi]$. $\varphi$ defines the position on the closed
curve. The topological phase $\alpha_{i,j}$ and the $D$-topological
matrix elements are both shown for the ``2$\times$H$^{35}$Cl + IR cavity
mode'' system. }
\label{fig:Berry_2} 
\end{figure}

In Fig. \ref{fig:Berry_2} topological results are presented for the
``2$\times$H$^{35}$Cl + IR cavity mode'' system in which, due to the
symmetry, only a second-order degeneracy point is formed (see on Fig.
\ref{fig:VPESs_2}). Consequently, if the two-state approximation
worked, one would have to obtain $2\pi$ for the value of the topological
phase $\alpha_{1,2}$. Instead, we obtained values of $\alpha_{1,2}=1.0878\pi$
and $\alpha_{2,3}=1.6789\pi$, which clearly show that the two-state
approximation does not work in the studied region of the CS. At this
point, one has to derive again the three-state ADT matrix elements
and the corresponding $\mathbf D$ matrix by employing the appropriate nonadiabatic
coupling terms. The obtained ($1$,$1$,$1$) values in the diagonal of
the $\mathbf D$ matrix clearly demonstrate that the three-state approximation
works perfectly and none of the vibrational wave functions change
sign. 

The above studies undoubtedly proved that the topological features
of VPES and electronic PES are very similar. Earlier results \cite{Baer_2002,Vertesi20053476,Baer_2006}
obtained for the latter situation can almost be transferred one by
one so as to understand the topological features of the VPESs. 
\blue{The topological properties of the VPESs imply that direct observables should exist, which prove the existence of the nonadiabatic effects among the VPESs. Some of these are presented in the following sections.}

\clearpage{}

\subsection{Light-dressed spectroscopy}

Another physical property which can be effected by CIs or LICIs is the spectrum. The spectroscopy of polaritonic systems is a valuable tool
for providing insight into their various physiochemical properties
\cite{spectra_of_vibDressedpolaritons_Zeb_ACSphotonics_2018,FRibeiro2018,Tamas_1a,Szidarovszky2020}, and nonadiabatic effects resulting from the strong
coupling of the electronic, vibrational, rotational, and photonic
degrees of freedom can cause a significant modification on the molecular absorption spectra.
In one possible theoretical formulation of light-dressed
spectroscopy \cite{Szidarovszky2019}, transitions between light-dressed
states are evaluated using first-order time-dependent perturbation
theory, that is, for the $T_{l\leftarrow k}$ transition amplitude
between the \textit{k}th and \textit{l}th light-dressed state $T_{l\leftarrow k}\propto\langle\Psi_{l}\vert\hat{\mathbf{d}}\mathbf{e}^{(p)}\vert\Psi_{k}\rangle\delta(E_{l}-E_{k}\pm\hbar\omega_{p})$
is assumed, where $\mathbf{e}^{(p)}$ and $\hbar\omega_{p}$ are the
polarization vector and photon energy of the probe pulse, respectively,
while $\vert\Psi_{i}\rangle$ and $E_{i}$ are the light-dressed wave
functions and energies, respectively. In this work the light-dressed
states (polaritonic states), $\vert\Psi_{i}\rangle$, have the form
shown in Eq. (\ref{eq:polaritonic_wave_function}), and the $T_{l\leftarrow k}$
transition amplitudes can be computed with formulae similar to that
in Eq. (\ref{eq:diatomic_hamiltonian_in_general_basis}). \blue{Note that within this framework the computed spectra represent the absorption of light running parallel to the cavity mirrors, in the medium, not that of which is transmitted through the mirrors.}

\begin{figure}[h!]
\centering \includegraphics[width=0.4\textwidth]{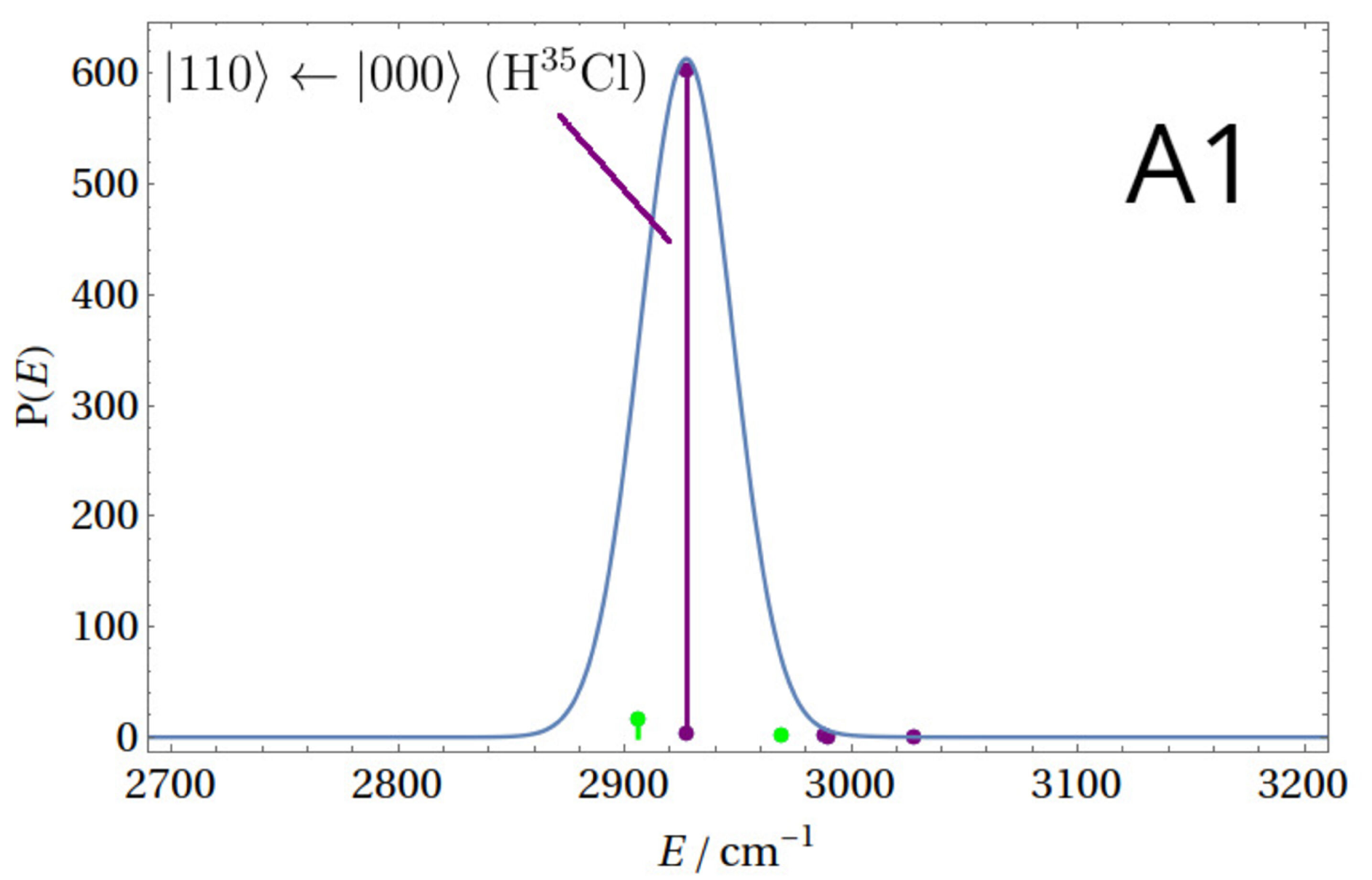}
\includegraphics[width=0.4\textwidth]{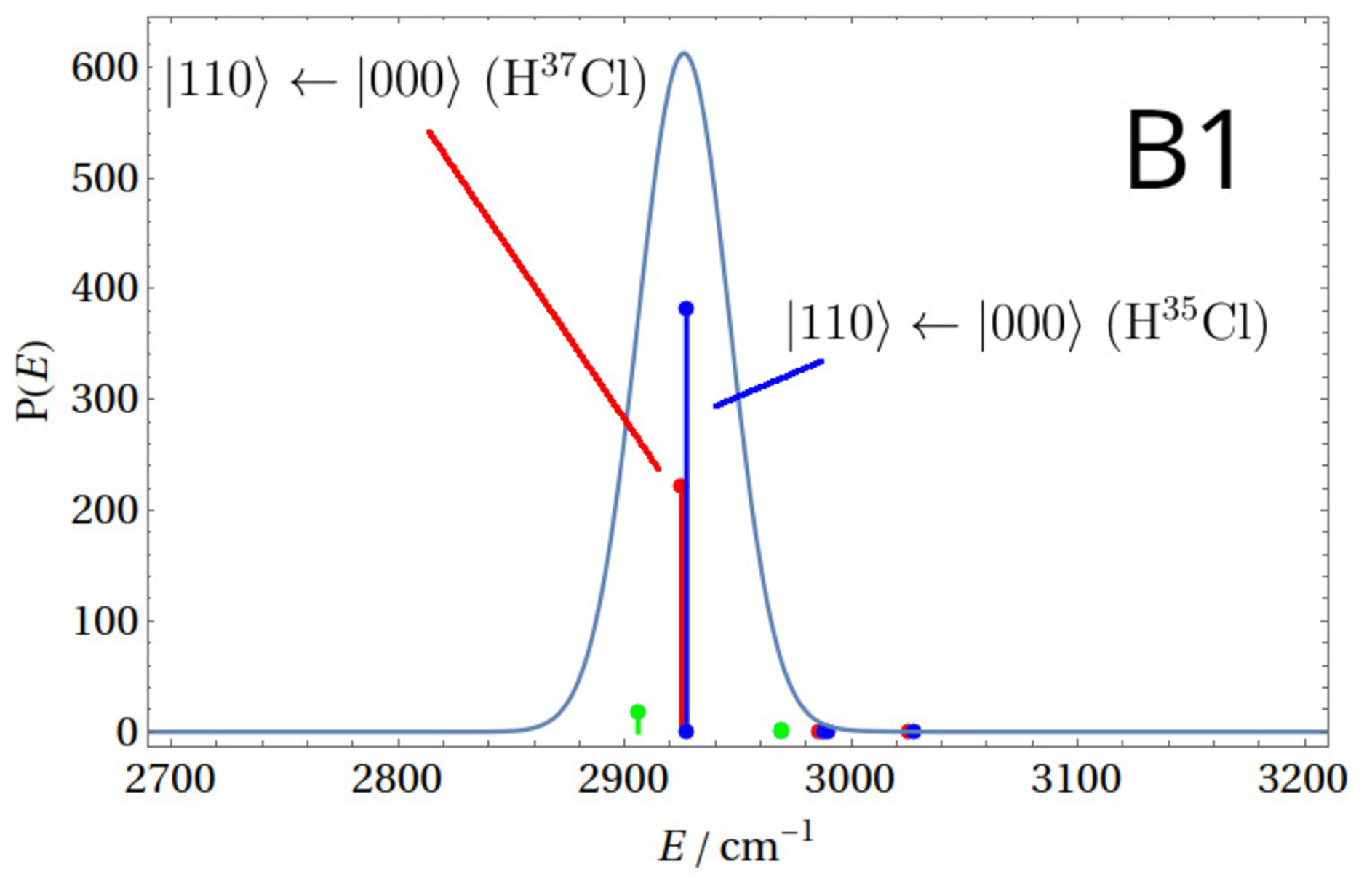}
\includegraphics[width=0.4\textwidth]{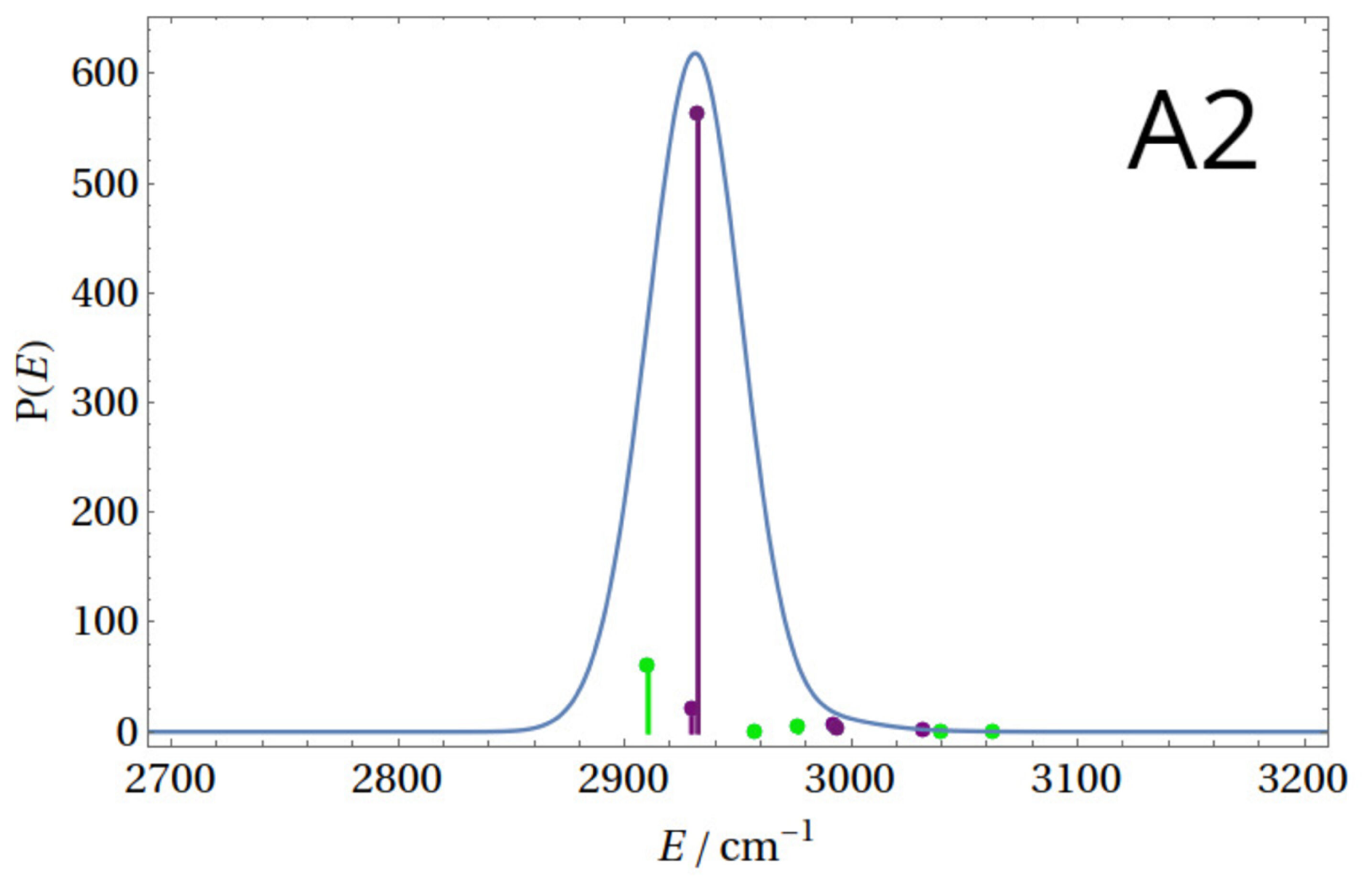}
\includegraphics[width=0.4\textwidth]{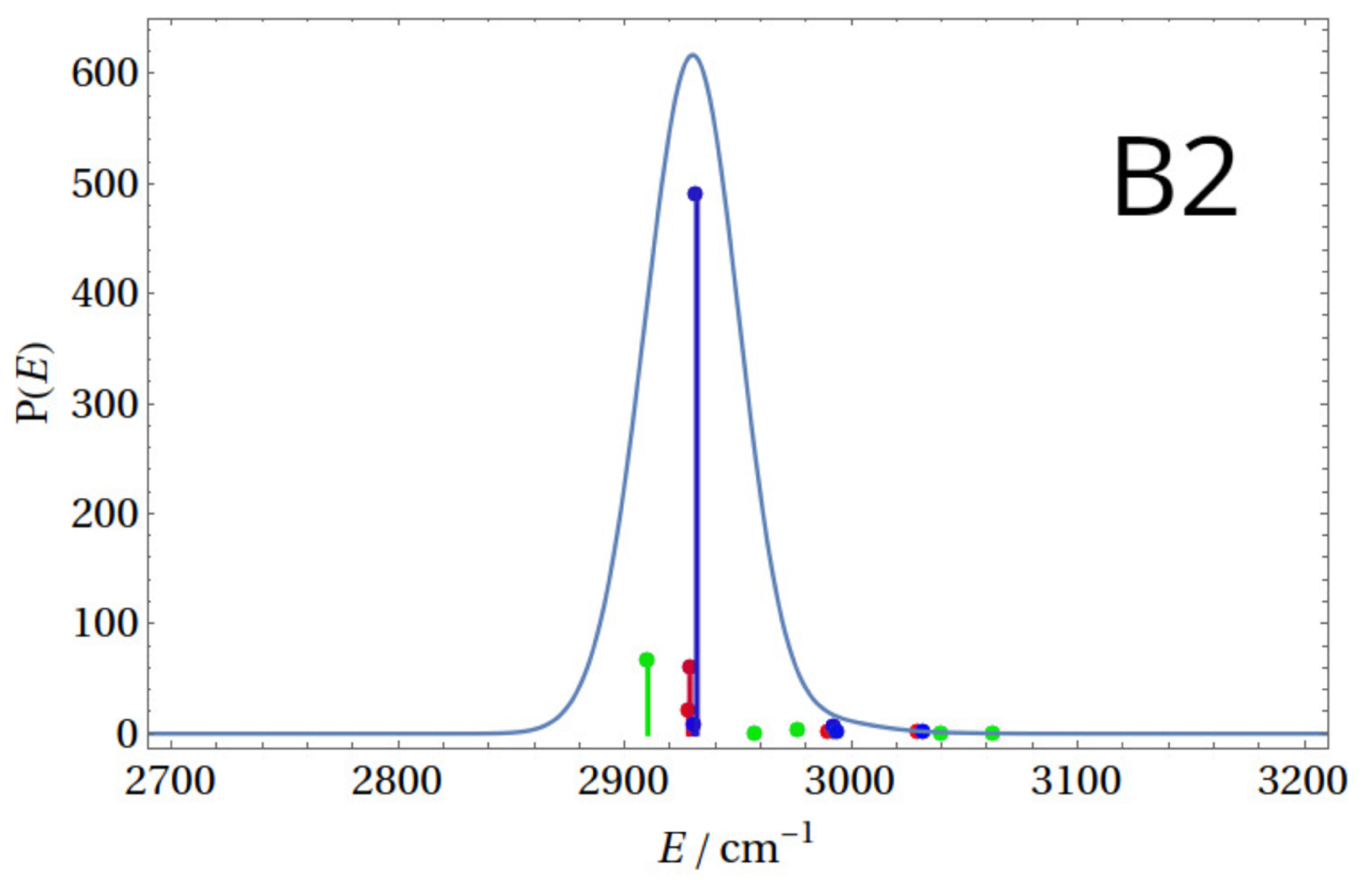}
\includegraphics[width=0.4\textwidth]{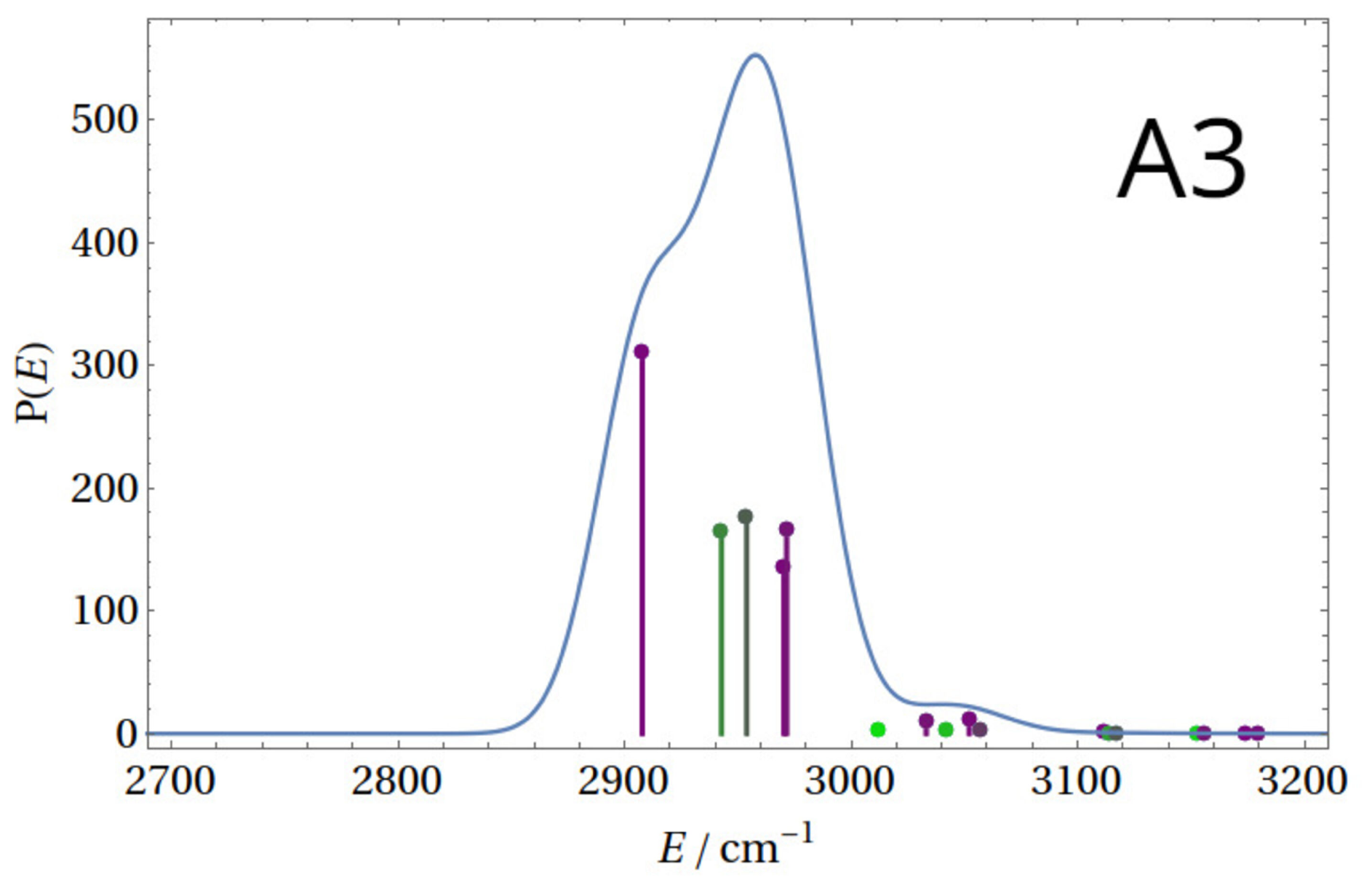}
\includegraphics[width=0.4\textwidth]{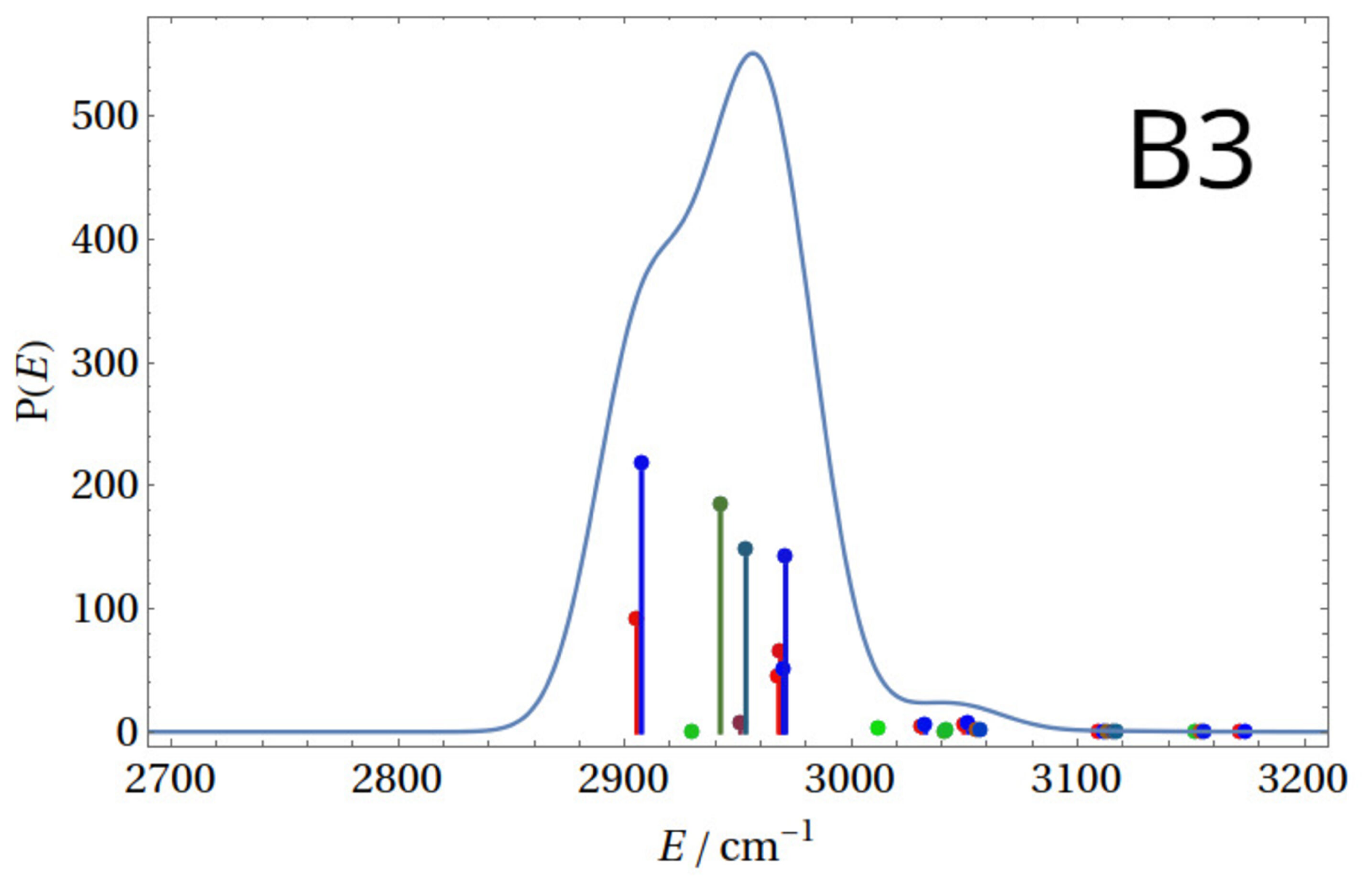}
\includegraphics[width=0.4\textwidth]{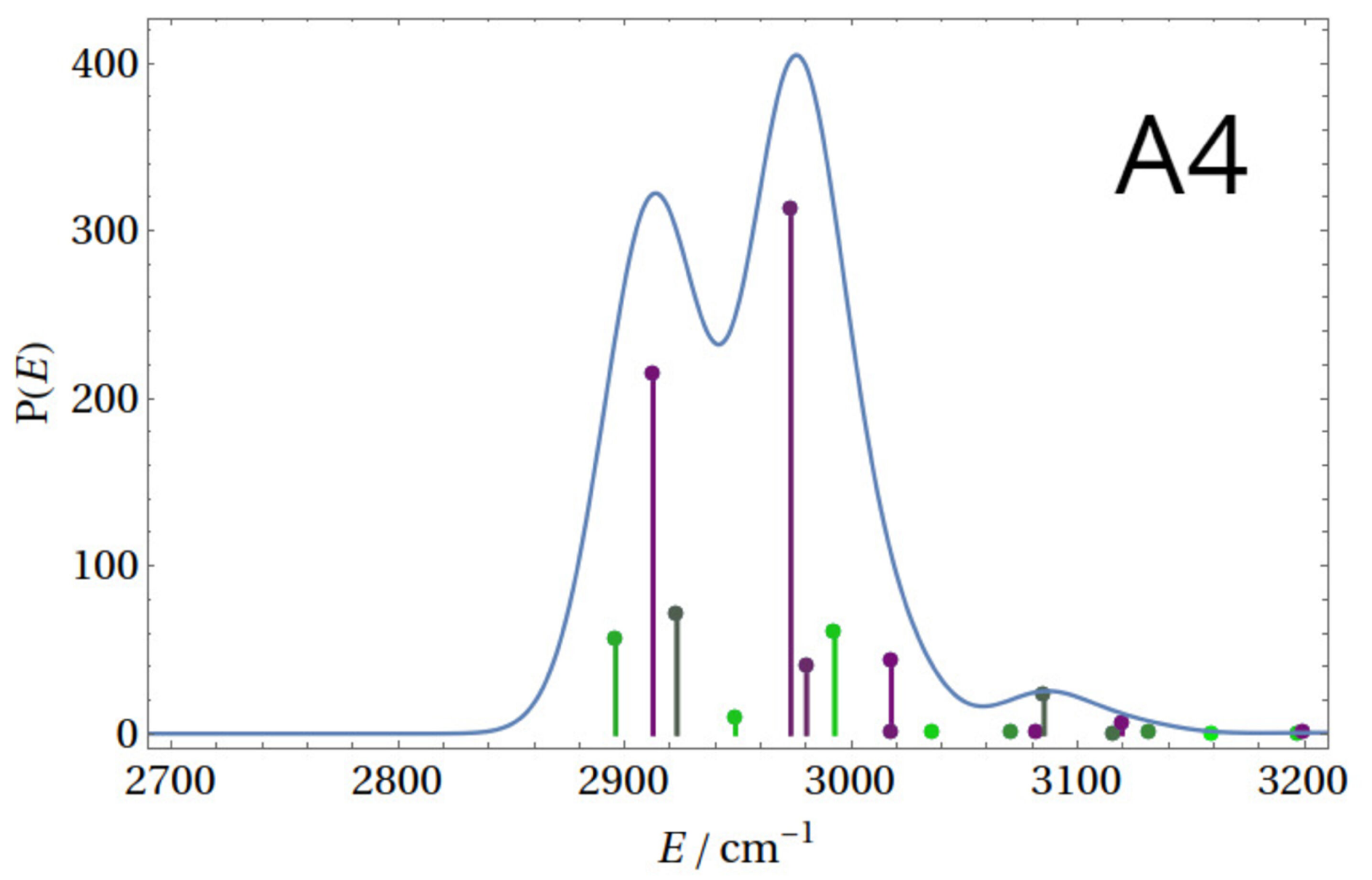}
\includegraphics[width=0.4\textwidth]{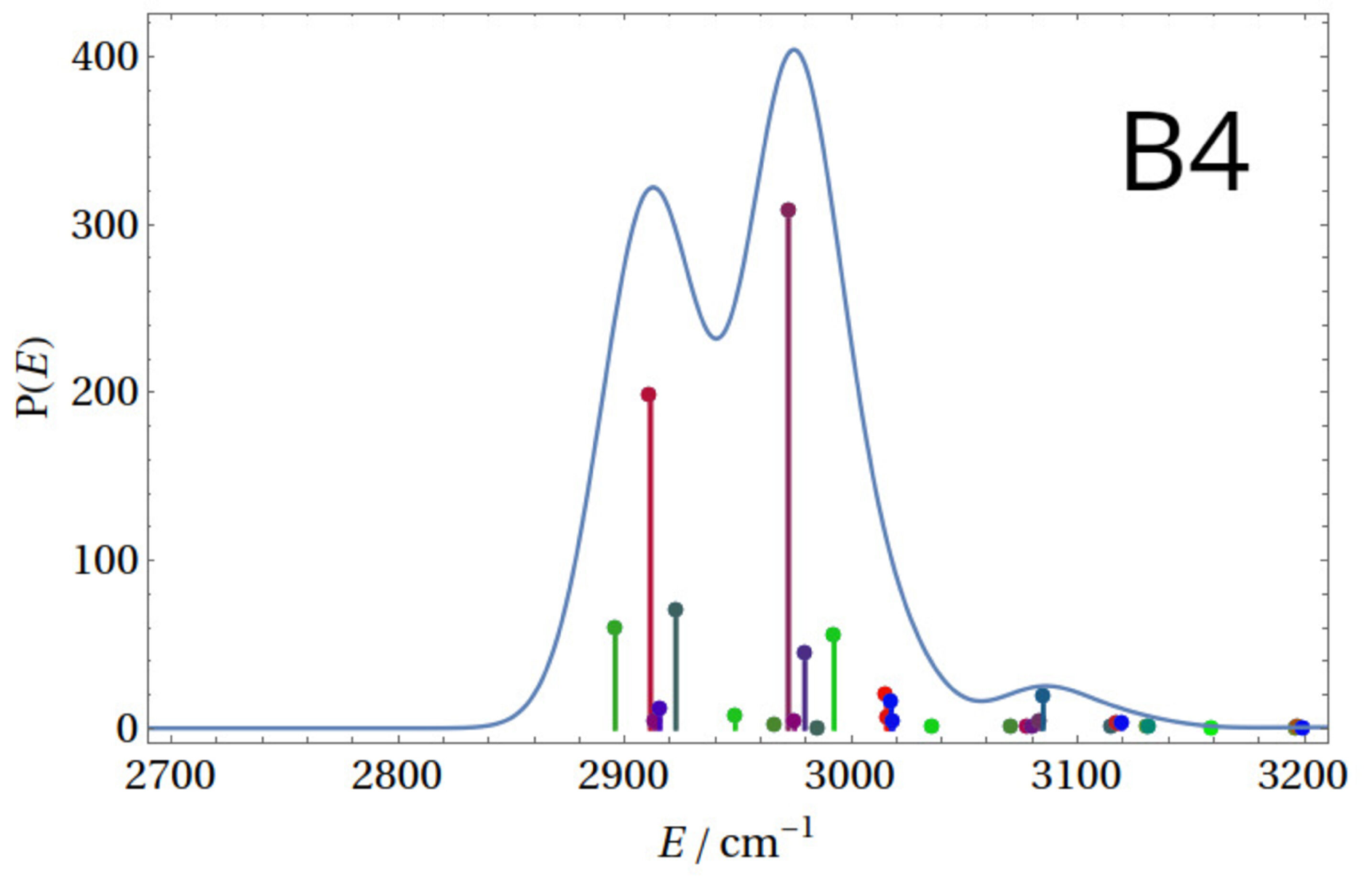}

\caption{Light-dressed spectrum of the ``2$\times$H$^{35}$Cl + IR cavity mode''
system (panels A1, A2, A3 and A4) and ``H$^{35}$Cl + H$^{37}$Cl + IR cavity
mode'' system (panels B1, B2, B3 and B4). See text for discussion on the peak assignment. Parameters are $E_{{\rm vib}}^{1,0}+E_{{\rm vib}}^{2,0}+\hbar\omega_{{\rm c}}=2904.5$
cm$^{-1}$, $E_{{\rm vib}}^{1,1}+E_{{\rm vib}}^{2,0}=2905.9$ cm$^{-1}$,
$E_{{\rm vib}}^{1,0}+E_{{\rm vib}}^{2,1}=2903.7$ cm$^{-1}$, and
the values of the coupling strength $\mu^{\rm (H^{35}Cl)}_{00}\sqrt{\hbar\omega_{{\rm c}}/(\varepsilon_{0}V)}$
equal 66.6, 133.2, 199.8,
and 266.4 cm$^{-1}$ in the different rows from top to bottom,
in order.}
\label{fig:light_dressed_spectra} 
\end{figure}

Figure \ref{fig:light_dressed_spectra} shows the IR absorption spectra
of H$^{35}$Cl molecules and the mixture of H$^{35}$Cl and H$^{37}$Cl
molecules, confined in IR cavities of different coupling strengths.
\blue{Blue, red, and green colors in the transition lines represent vibrational excitation in H$^{35}$Cl, vibrational excitation in the other H$^{35}$Cl (left panels) or H$^{37}$Cl (right panels), and photonic excitation, respectively.
The color of each transition line is obtained by mixing the above three colors according to the weight of the different types of basis functions in the final polaritonic states of the transition.}
The panel (A1) in Fig. \ref{fig:light_dressed_spectra}
demonstrates that at the weakest coupling strength presented in this
work, the light-dressed spectrum of two confined H$^{35}$Cl molecules
shows one dominant and several minor peaks, arising from the splitting of the single
$\vert v=1,\,J=1,\,M=0\rangle\leftarrow\vert v=0,\,J=0,\,M=0\rangle$
field-free transition. The strongest peak is split into two in panel
(B1) of Fig. \ref{fig:light_dressed_spectra}. \blue{The small energy splitting reflects the slight differences in the rovibrational energies of H$^{35}$Cl and H$^{37}$Cl, while the significant difference in the peak heights reflects the different VPES landscape and related nonadiabatic effects (under field-free conditions the height of the split peak is nearly identical).
The small green peaks represent excitation in the photonic mode (photon number increased by one), which appears in the spectrum due to the contamination of the photonic mode with  molecular excited (ro)vibrational states, having non-zero transition amplitude.
No significant difference can be seen in the low-resolution envelopes of the two panels in the first row.

The light dressed spectra at larger cavity coupling strengths demonstrate that the individual peak positions and peak heights can be slightly different for the confined 2$\times$H$^{35}$Cl and H$^{35}$Cl+H$^{37}$Cl systems, but no significant differences arise in their spectrum envelope. 
Due to the strong mixing between the different molecular excitations and the photonic mode, assigning the peaks at larger cavity coupling strengths becomes complicated and less informative. 
Nonetheless, the peak colors represent the character of the different transitions to some extent.
Panels (B2),(B3) and (B4) of Fig. \ref{fig:light_dressed_spectra} show that some peaks are dominated by a H$^{35}$Cl, H$^{37}$Cl or photonic transition (peaks with mostly red, blue or green color), but other peaks show significant mixing between the rovibrational modes of the HCl molecules and the photonic mode (purple, dark green or brown peaks).
Despite the subtle differences in the spectra of the ``2$\times$H$^{35}$Cl + cavity mode\char`\"{} and ``H$^{35}$Cl + H$^{37}$Cl cavity mode\char`\"{} systems, some of which can be attributed to the nonadiabatic couplings between the VPESs, the low-resolution spectrum envelopes remain nearly identical, thus the IR spectrum doesn't seem to be the most efficient tool to highlight the nonadiabatic effects in this case.}

\clearpage{}

\subsection{Laser-induced dynamics}

\blue{An additional approach to highlight nonadiabatic couplings in molecular systems is to investigate their laser-induced dynamics. In this work such simulations were carried out by directly solving the time-dependent Schr{\"o}dinger equation in the diabatic direct-product basis representation, assuming $T=0$ K. At room temperature, due to the non-zero populations in $J>0$ rotational levels, the numerical values for the expectation values shown below would vary, but the qualitative conclusions should not change.}

As already demonstrated for the vibrational dynamics on electronic
PESs \cite{LICI1}, an efficient way to pinpoint nonadiabatic effects
is to monitor the populations on the adiabatic surfaces during the
dynamics. Here we perform similar studies for the case of
adiabatic vibrational polaritons. In the case of VPESs, the population
on the $a$th adiabatic state is evaluated as 
\begin{equation}
p^{(a)}=\vert\hat{P}^{(a)}\vert\Psi(t)\rangle\vert^{2},
\end{equation}
where $\vert\Psi(t)\rangle$ is the time-dependent wave function expanded
in terms of the direct-product diabatic states, see Eq. (\ref{eq:direct_product_basis}),
and $\hat{P}^{(a)}=\vert\psi^{(a)}\rangle\langle\psi^{(a)}\vert$
is a projector onto the $a$th adiabatic state, \textit{i.e.}, the
$a$th eigenstate of the matrix in Eq. (\ref{eq:VPES_hamiltonian}).
Note that 
\begin{equation}
\vert\psi^{a}\rangle=\sum_{N,v_{1},v_{2}}{C_{N,v_{1},v_{2}}^{(a)}(\theta_{1},\theta_{2})\vert N\rangle\vert\Psi_{{\rm vib}}^{1,v_{1}}\rangle\vert\Psi_{{\rm vib}}^{2,v_{2}}\rangle},
\end{equation}
that is, the $C_{N,v_{1},v_{2}}^{(a)}(\theta_{1},\theta_{2})$ expansion
coefficients depend on the rotational coordinates, $\theta_{1}$ and
$\theta_{2}$, as predictable from Figs. \ref{fig:VPESs} and \ref{fig:VPESs_2}.

\begin{figure}[h!]
\centering \includegraphics[width=0.95\textwidth]{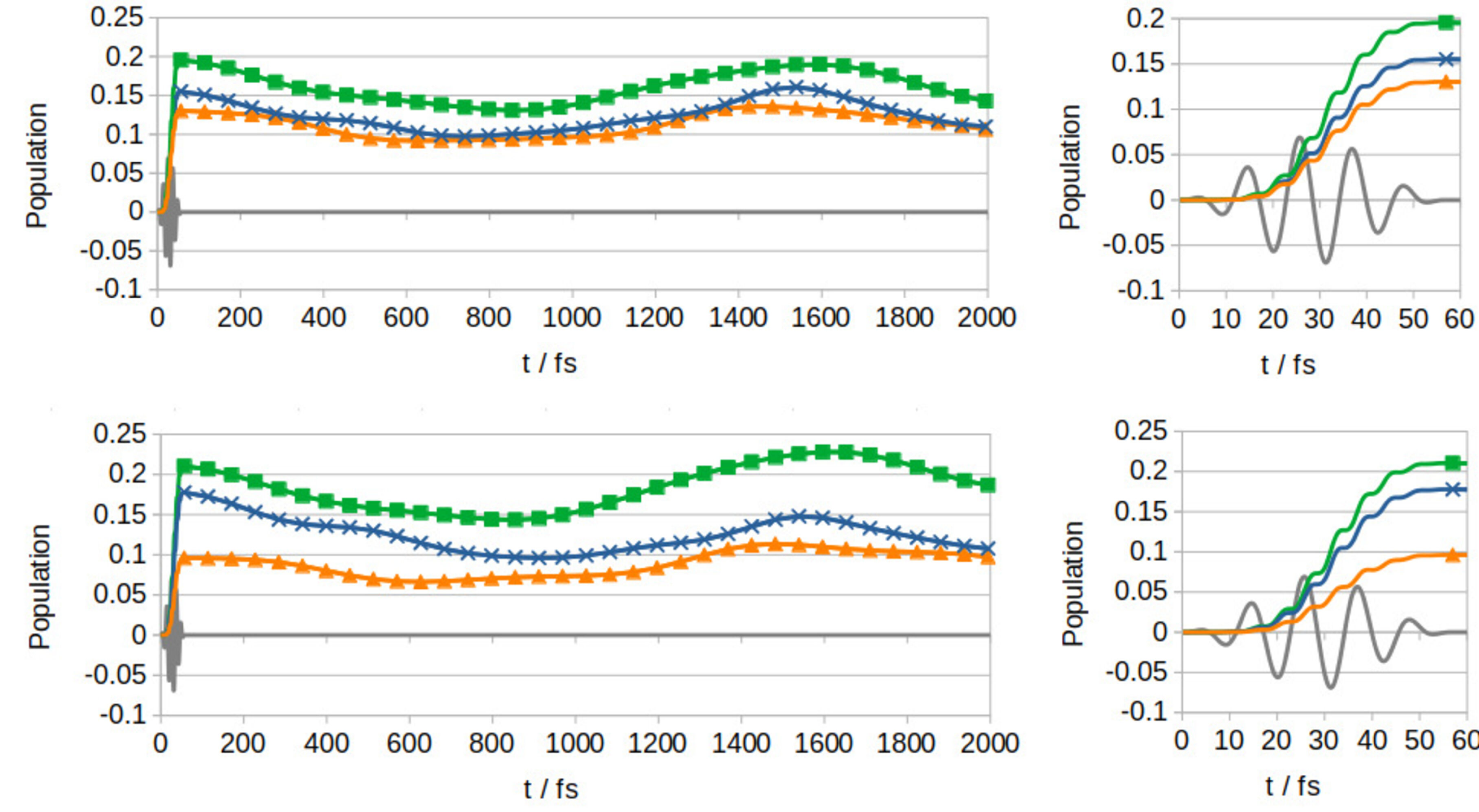}

\caption{Adiabatic populations in the laser-induced dynamics of the ``2$\times$H$^{35}$Cl + IR cavity mode'' system (upper panels) and ``H$^{35}$Cl + H$^{37}$Cl + IR cavity mode'' system (lower panels). Squares, crosses,
and triangles stand for the upper, middle, and lower adiabatic polaritonic
VPESs, respectively, color coded according to Figs. \ref{fig:VPESs}
and \ref{fig:VPESs_2}. 
%The dashed line shows the ground state VPES population, and
The continuous line represents the electric field
of the laser pulse. Parameters are $E_{{\rm vib}}^{1,0}+E_{{\rm vib}}^{2,0}+\hbar\omega_{{\rm c}}=2904.5$
cm$^{-1}$, $E_{{\rm vib}}^{1,1}+E_{{\rm vib}}^{2,0}=2905.9$ cm$^{-1}$,
$E_{{\rm vib}}^{1,0}+E_{{\rm vib}}^{2,1}=2903.7$ cm$^{-1}$, and
$\mu^{\rm (H^{35}Cl)}\sqrt{\hbar\omega_{{\rm c}}/(\varepsilon_{0}V)}=133.2$
cm$^{-1}$. The pump laser has a field strength of $0.1/\sqrt{2}$ a.u.,
and its central wavenumber equals the field-free $\vert11\rangle\leftarrow\vert00\rangle$
rovibrational transition of H$^{35}$Cl (2926.1 cm$^{-1}$).}
\label{fig:2RR_adpops_in_dynamics} 
\end{figure}

\blue{
Fig. \ref{fig:2RR_adpops_in_dynamics} shows the temporal evolution
of the adiabatic populations for the ``2$\times$H$^{35}$Cl + IR cavity
mode'' system and the ``H$^{35}$Cl + H$^{37}$Cl + IR cavity mode''
system, when a short IR laser pulse, tuned to the field-free vibrational
fundamental of H$^{35}$Cl (2926.1 cm$^{-1}$), is used to excite
these systems from their ground states. 
After 60 fs, when the laser field has subsided, the population of the ground adiabatic state is 0.52 for both systems, indicating that the degree of laser-induced molecular vibrational excitation is identical for the two systems. 
For both the ``2$\times$H$^{35}$Cl + IR cavity mode'' and ``H$^{35}$Cl + H$^{37}$Cl + IR cavity mode'' systems, shown in the the upper and lower panels of Fig. \ref{fig:2RR_adpops_in_dynamics}, respectively, the laser populates all three excited adiabatic states. 
However, the population is distributed among the three excited adiabatic states in a different manner for the two systems. 
Somewhat surprisingly, the middle polaritons, which one would intuitively address as dark states, are also populated by the light field for both systems.
This is due to the permanent dipole coupling with the ground adiabatic state (see first row and column of second matrix in Eq. (\ref{eq:VPES_hamiltonian})), test simulations without this coupling show that the middle polaritons are indeed almost ``dark'', and are much less populated.
In either case, due to the strong nonadiabatic couplings between the VPESs, rapid, hundred-femtosecond timescale population transfer proceeds among all three adiabatic states following the excitation by the laser field. 
}
Overall, Fig. \ref{fig:2RR_adpops_in_dynamics} demonstrates that when molecular rotations are feasable, strong nonadiabatic
couplings influence the dynamics of multimolecule (ro)vibrational polaritons. 
Furthermore, the nonadiabatic dynamics shown in Fig. \ref{fig:2RR_adpops_in_dynamics} verify that breaking the molecular permutational symmetry, in this case by introducing a different Cl isotope, the polariton dynamics can deviate substantially.

\begin{figure}[h!]
\centering \includegraphics[width=0.95\textwidth]{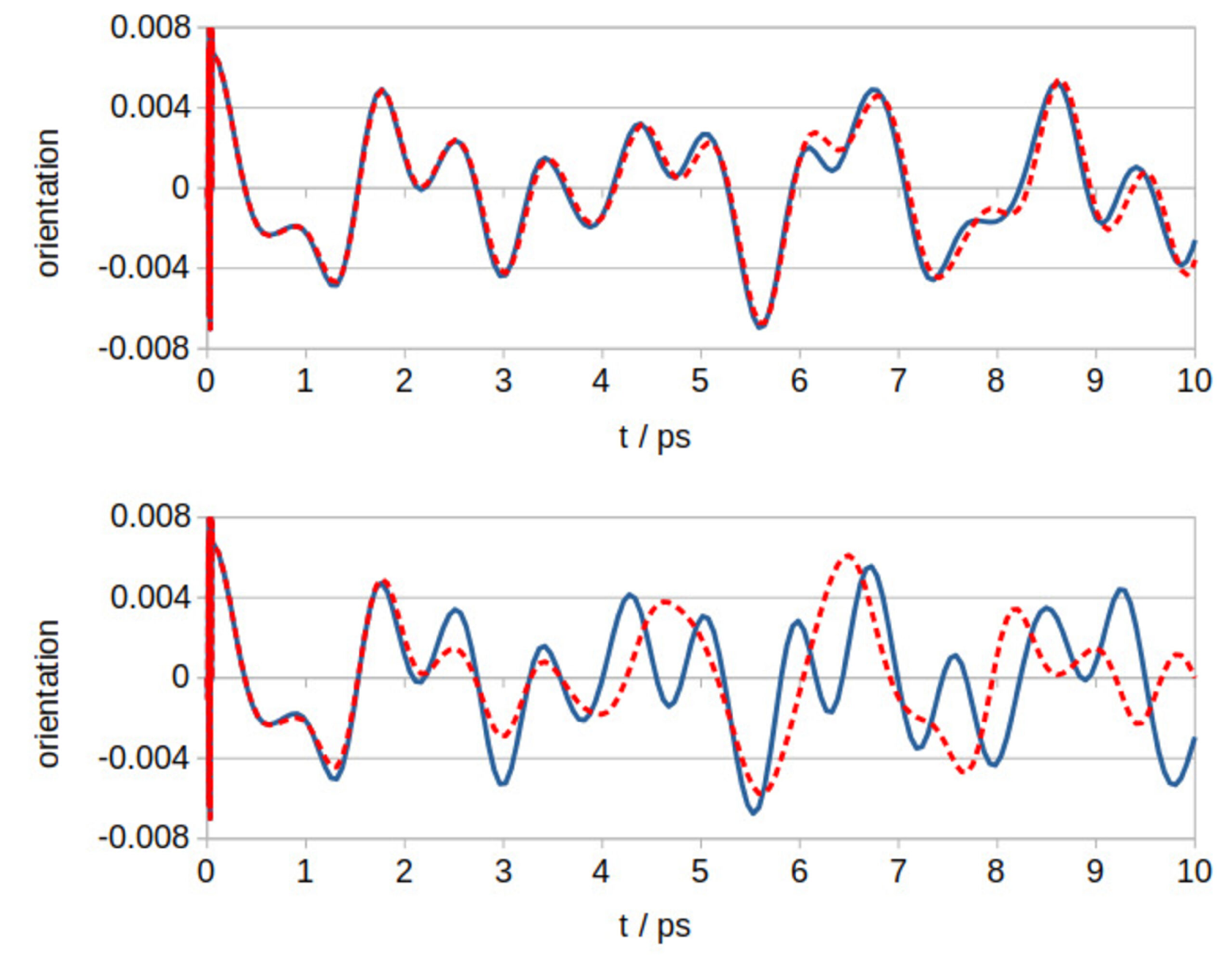}

\caption{Upper panel: the laser-induced orientation, \textit{i.e.}, the $\langle {\rm cos}(\theta)\rangle$ expectation value of a confined H$^{35}$Cl molecule (blue continuous line) or a confined H$^{37}$Cl molecule (red dashed line), when sharing the cavity with an identical isotopologue, \textit{i.e.}, with a H$^{35}$Cl and a H$^{37}$Cl, respectively. Lower panel: the laser-induced orientation of H$^{35}$Cl (blue continuous line) and H$^{37}$Cl (red dashed line) in the ``H$^{35}$Cl + H$^{37}$Cl + IR cavity mode'' system. The inverse rotational constant in time units is 0.509 ps and 0.510 ps for H$^{35}$Cl and H$^{37}$Cl, respectively.  $\mu^{\rm (H^{35}Cl)}\sqrt{\hbar\omega_{{\rm c}}/(\varepsilon_{0}V)}=133.2$
cm$^{-1}$, all other parameters are the same as in Fig. \ref{fig:2RR_adpops_in_dynamics}.}
\label{fig:2RR_orientation_dynamics} 
\end{figure}

As an alternative approach to probe the presence of nonadiabatic couplings in the laser-induced dynamics of our rovibrational polaritonic test systems, the orientation, \textit{i.e.}, the $\langle {\rm cos}(\theta) \rangle$ expectation value, was evaluated for a single H$^{35}$Cl (H$^{37}$Cl) molecule, when confined in an IR cavity with another H$^{35}$Cl (H$^{37}$Cl) molecule or a H$^{37}$Cl (H$^{35}$Cl) isotopologue. 
\blue{The numerical results are shown for the same and mixed isotopologue cases in the upper and lower panels of Figure \ref{fig:2RR_orientation_dynamics}, respectively.} 
\blue{The upper panel of Figure \ref{fig:2RR_orientation_dynamics} demonstrates that the isotope effect causes only minor differences in the orientation dinamics, the curves for H$^{35}$Cl and H$^{37}$Cl and nearly identical. However, the lower panel of Figure \ref{fig:2RR_orientation_dynamics} shows that on the timescale of the rotational dynamics of HCl (picoseconds), the orientation curves deviate substantially when the two isotopologues are mixed in the cavity.}
Although the exact value of the numerical results presented here should be treated with caution, because cavity loss was neglected in this work, such a dependence of the orientation on the partner molecule is a clear fingerprint of molecular distinguishability on the VPESs (and corresponding nonadiabatic couplings) and, therefore, on the rotational dynamics.

\clearpage{}

\section{Summary and conclusions}

A general framework for computing the rovibrational polaritons of several molecules %ar ensembles
in a lossless cavity was presented.
Based on this formulation, the concept of vibrational potential energy surfaces (VPES) was introduced, which provide effective potential energy surfaces for the rotational motion of the confined molecules.
For the test system of two rovibrating HCl molecules interacting with a single lossless infrared cavity mode, degeneracies were identified between the VPESs: light-induced conical intersections (LICI) were found when two different isotopologoues, H$^{35}$Cl and H$^{37}$Cl, are in the cavity, and a second-order degeneracy was found in the case of two same isotopologoues.
The degeneracies between the VPESs were characterized based on their topological properties, and their nonadiabatic fingerprints were identified in the light-dressed spectra and the laser-induced dynamical properties of the investigated rovibrational polaritons. 

The presented results clearly show that breaking molecular permutational symmetry due to different isotopes can play an important role in rovibrational polaritonic properties; the polaritonic surfaces, nonadiabatic couplings, and related spectral, topological, and dynamic properties can be influenced substantially. This implies that the natural abundance of different isotopologoues should be accounted for if one aims for a realistic simulation on confined molecular ensembles.

%$\vert 110\rangle\leftarrow\vert 000\rangle$ (H$^{35}$Cl)

%$\vert 110\rangle\leftarrow\vert 000\rangle$ (H$^{37}$Cl)

\section{Acknowledgement}

This research was supported by the EU-funded Hungarian grant EFOP-3.6.2-16-2017-00005 as well as by the János Bolyai Research Scholarship of the Hungarian Academy of Sciences and by the ÚNKP-20-5 New National Excellence Program of the Ministry for Innovation and Technology from the source of the National Research, Development and Innovation Fund.
The authors are grateful to NKFIH for support (Grants No. PD124623, K128396 and FK134291).

\section*{Data availability}
The data that support the findings of this study are available from the corresponding author upon reasonable request.

\clearpage{}

\bibliography{main}

\begin{thebibliography}{75}%
\makeatletter
\providecommand \@ifxundefined [1]{%
 \@ifx{#1\undefined}
}%
\providecommand \@ifnum [1]{%
 \ifnum #1\expandafter \@firstoftwo
 \else \expandafter \@secondoftwo
 \fi
}%
\providecommand \@ifx [1]{%
 \ifx #1\expandafter \@firstoftwo
 \else \expandafter \@secondoftwo
 \fi
}%
\providecommand \natexlab [1]{#1}%
\providecommand \enquote  [1]{``#1''}%
\providecommand \bibnamefont  [1]{#1}%
\providecommand \bibfnamefont [1]{#1}%
\providecommand \citenamefont [1]{#1}%
\providecommand \href@noop [0]{\@secondoftwo}%
\providecommand \href [0]{\begingroup \@sanitize@url \@href}%
\providecommand \@href[1]{\@@startlink{#1}\@@href}%
\providecommand \@@href[1]{\endgroup#1\@@endlink}%
\providecommand \@sanitize@url [0]{\catcode `\\12\catcode `\$12\catcode
  `\&12\catcode `\#12\catcode `\^12\catcode `\_12\catcode `\%12\relax}%
\providecommand \@@startlink[1]{}%
\providecommand \@@endlink[0]{}%
\providecommand \url  [0]{\begingroup\@sanitize@url \@url }%
\providecommand \@url [1]{\endgroup\@href {#1}{\urlprefix }}%
\providecommand \urlprefix  [0]{URL }%
\providecommand \Eprint [0]{\href }%
\providecommand \doibase [0]{http://dx.doi.org/}%
\providecommand \selectlanguage [0]{\@gobble}%
\providecommand \bibinfo  [0]{\@secondoftwo}%
\providecommand \bibfield  [0]{\@secondoftwo}%
\providecommand \translation [1]{[#1]}%
\providecommand \BibitemOpen [0]{}%
\providecommand \bibitemStop [0]{}%
\providecommand \bibitemNoStop [0]{.\EOS\space}%
\providecommand \EOS [0]{\spacefactor3000\relax}%
\providecommand \BibitemShut  [1]{\csname bibitem#1\endcsname}%
\let\auto@bib@innerbib\@empty
%</preamble>
\bibitem [{\citenamefont {Miller}\ \emph {et~al.}(2005)\citenamefont {Miller},
  \citenamefont {Northup}, \citenamefont {Birnbaum}, \citenamefont {Boca},
  \citenamefont {Boozer},\ and\ \citenamefont {Kimble}}]{Miller_2005}%
  \BibitemOpen
  \bibfield  {author} {\bibinfo {author} {\bibfnamefont {R.}~\bibnamefont
  {Miller}}, \bibinfo {author} {\bibfnamefont {T.~E.}\ \bibnamefont {Northup}},
  \bibinfo {author} {\bibfnamefont {K.~M.}\ \bibnamefont {Birnbaum}}, \bibinfo
  {author} {\bibfnamefont {A.}~\bibnamefont {Boca}}, \bibinfo {author}
  {\bibfnamefont {A.~D.}\ \bibnamefont {Boozer}}, \ and\ \bibinfo {author}
  {\bibfnamefont {H.~J.}\ \bibnamefont {Kimble}},\ }\href {\doibase
  10.1088/0953-4075/38/9/007} {\bibfield  {journal} {\bibinfo  {journal}
  {Journal of Physics B: Atomic, Molecular and Optical Physics}\ }\textbf
  {\bibinfo {volume} {38}},\ \bibinfo {pages} {S551} (\bibinfo {year}
  {2005})}\BibitemShut {NoStop}%
\bibitem [{\citenamefont {Walls}\ and\ \citenamefont
  {Milburn}(2008)}]{Walls_2008}%
  \BibitemOpen
  \bibfield  {author} {\bibinfo {author} {\bibfnamefont {D.~F.}\ \bibnamefont
  {Walls}}\ and\ \bibinfo {author} {\bibfnamefont {G.~J.}\ \bibnamefont
  {Milburn}},\ }\href@noop {} {\emph {\bibinfo {title} {Quantum Optics}}},\
  Berlin Heidelberg\ (\bibinfo  {publisher} {Springer},\ \bibinfo {year}
  {2008})\BibitemShut {NoStop}%
\bibitem [{\citenamefont {Ruggenthaler}\ \emph {et~al.}(2018)\citenamefont
  {Ruggenthaler}, \citenamefont {Tancogne-Dejean}, \citenamefont {Flick},
  \citenamefont {Appel},\ and\ \citenamefont {Rubio}}]{Ruggenthaler_2018}%
  \BibitemOpen
  \bibfield  {author} {\bibinfo {author} {\bibfnamefont {M.}~\bibnamefont
  {Ruggenthaler}}, \bibinfo {author} {\bibfnamefont {N.}~\bibnamefont
  {Tancogne-Dejean}}, \bibinfo {author} {\bibfnamefont {J.}~\bibnamefont
  {Flick}}, \bibinfo {author} {\bibfnamefont {H.}~\bibnamefont {Appel}}, \ and\
  \bibinfo {author} {\bibfnamefont {A.}~\bibnamefont {Rubio}},\ }\href
  {http://dx.doi.org/10.1038/s41570-018-0118} {\bibfield  {journal} {\bibinfo
  {journal} {Nat. Rev. Chem.}\ }\textbf {\bibinfo {volume} {2}},\ \bibinfo
  {pages} {0118} (\bibinfo {year} {2018})}\BibitemShut {NoStop}%
\bibitem [{\citenamefont {Herrera}\ and\ \citenamefont
  {Owrutsky}(2020)}]{Herrera_2020}%
  \BibitemOpen
  \bibfield  {author} {\bibinfo {author} {\bibfnamefont {F.}~\bibnamefont
  {Herrera}}\ and\ \bibinfo {author} {\bibfnamefont {J.}~\bibnamefont
  {Owrutsky}},\ }\href {\doibase 10.1063/1.5136320} {\bibfield  {journal}
  {\bibinfo  {journal} {The Journal of Chemical Physics}\ }\textbf {\bibinfo
  {volume} {152}},\ \bibinfo {pages} {100902} (\bibinfo {year}
  {2020})}\BibitemShut {NoStop}%
\bibitem [{\citenamefont {Hutchison}\ \emph {et~al.}(2012)\citenamefont
  {Hutchison}, \citenamefont {Schwartz}, \citenamefont {Genet}, \citenamefont
  {Devaux},\ and\ \citenamefont {Ebbesen}}]{cavity_exp_Hutchison_AngChem_2012}%
  \BibitemOpen
  \bibfield  {author} {\bibinfo {author} {\bibfnamefont {J.~A.}\ \bibnamefont
  {Hutchison}}, \bibinfo {author} {\bibfnamefont {T.}~\bibnamefont {Schwartz}},
  \bibinfo {author} {\bibfnamefont {C.}~\bibnamefont {Genet}}, \bibinfo
  {author} {\bibfnamefont {E.}~\bibnamefont {Devaux}}, \ and\ \bibinfo {author}
  {\bibfnamefont {T.~W.}\ \bibnamefont {Ebbesen}},\ }\href {\doibase
  10.1002/anie.201107033} {\bibfield  {journal} {\bibinfo  {journal} {Angew.
  Chem. Int. Ed.}\ }\textbf {\bibinfo {volume} {51}},\ \bibinfo {pages} {1592}
  (\bibinfo {year} {2012})}\BibitemShut {NoStop}%
\bibitem [{\citenamefont {Ebbesen}(2016)}]{cavity_Ebbesen_AccChemRes_2016}%
  \BibitemOpen
  \bibfield  {author} {\bibinfo {author} {\bibfnamefont {T.~W.}\ \bibnamefont
  {Ebbesen}},\ }\href {\doibase 10.1021/acs.accounts.6b00295} {\bibfield
  {journal} {\bibinfo  {journal} {Acc. Chem. Res.}\ }\textbf {\bibinfo {volume}
  {49}},\ \bibinfo {pages} {2403} (\bibinfo {year} {2016})}\BibitemShut
  {NoStop}%
\bibitem [{\citenamefont {Hertzog}\ \emph {et~al.}(2019)\citenamefont
  {Hertzog}, \citenamefont {Wang}, \citenamefont {Mony},\ and\ \citenamefont
  {B\"{o}rjesson}}]{Hertzog2019}%
  \BibitemOpen
  \bibfield  {author} {\bibinfo {author} {\bibfnamefont {M.}~\bibnamefont
  {Hertzog}}, \bibinfo {author} {\bibfnamefont {M.}~\bibnamefont {Wang}},
  \bibinfo {author} {\bibfnamefont {J.}~\bibnamefont {Mony}}, \ and\ \bibinfo
  {author} {\bibfnamefont {K.}~\bibnamefont {B\"{o}rjesson}},\ }\href {\doibase
  10.1039/c8cs00193f} {\bibfield  {journal} {\bibinfo  {journal} {Chemical
  Society Reviews}\ }\textbf {\bibinfo {volume} {48}},\ \bibinfo {pages} {937}
  (\bibinfo {year} {2019})}\BibitemShut {NoStop}%
\bibitem [{\citenamefont {Galego}\ \emph {et~al.}(2015)\citenamefont {Galego},
  \citenamefont {Garcia-Vidal},\ and\ \citenamefont
  {Feist}}]{cavity_Galego_PRX_2015}%
  \BibitemOpen
  \bibfield  {author} {\bibinfo {author} {\bibfnamefont {J.}~\bibnamefont
  {Galego}}, \bibinfo {author} {\bibfnamefont {F.~J.}\ \bibnamefont
  {Garcia-Vidal}}, \ and\ \bibinfo {author} {\bibfnamefont {J.}~\bibnamefont
  {Feist}},\ }\href {\doibase 10.1103/PhysRevX.5.041022} {\bibfield  {journal}
  {\bibinfo  {journal} {Phys. Rev. X}\ }\textbf {\bibinfo {volume} {5}},\
  \bibinfo {pages} {1} (\bibinfo {year} {2015})}\BibitemShut {NoStop}%
\bibitem [{\citenamefont {Galego}\ \emph {et~al.}(2016)\citenamefont {Galego},
  \citenamefont {Garcia-Vidal},\ and\ \citenamefont
  {Feist}}]{cavity_Galego_NatCommun_2016}%
  \BibitemOpen
  \bibfield  {author} {\bibinfo {author} {\bibfnamefont {J.}~\bibnamefont
  {Galego}}, \bibinfo {author} {\bibfnamefont {F.~J.}\ \bibnamefont
  {Garcia-Vidal}}, \ and\ \bibinfo {author} {\bibfnamefont {J.}~\bibnamefont
  {Feist}},\ }\href {\doibase 10.1038/ncomms13841} {\bibfield  {journal}
  {\bibinfo  {journal} {Nat. Commun.}\ }\textbf {\bibinfo {volume} {7}},\
  \bibinfo {pages} {13841} (\bibinfo {year} {2016})}\BibitemShut {NoStop}%
\bibitem [{\citenamefont {Feist}\ \emph {et~al.}(2018)\citenamefont {Feist},
  \citenamefont {Galego},\ and\ \citenamefont
  {Garcia-Vidal}}]{cavity_Feist_ACSphotonics_2018}%
  \BibitemOpen
  \bibfield  {author} {\bibinfo {author} {\bibfnamefont {J.}~\bibnamefont
  {Feist}}, \bibinfo {author} {\bibfnamefont {J.}~\bibnamefont {Galego}}, \
  and\ \bibinfo {author} {\bibfnamefont {F.~J.}\ \bibnamefont {Garcia-Vidal}},\
  }\href {\doibase 10.1021/acsphotonics.7b00680} {\bibfield  {journal}
  {\bibinfo  {journal} {ACS Photonics}\ }\textbf {\bibinfo {volume} {5}},\
  \bibinfo {pages} {205} (\bibinfo {year} {2018})}\BibitemShut {NoStop}%
\bibitem [{\citenamefont {Sch\"{a}fer}\ \emph {et~al.}(2018)\citenamefont
  {Sch\"{a}fer}, \citenamefont {Ruggenthaler},\ and\ \citenamefont
  {Rubio}}]{Schafer_2018}%
  \BibitemOpen
  \bibfield  {author} {\bibinfo {author} {\bibfnamefont {C.}~\bibnamefont
  {Sch\"{a}fer}}, \bibinfo {author} {\bibfnamefont {M.}~\bibnamefont
  {Ruggenthaler}}, \ and\ \bibinfo {author} {\bibfnamefont {A.}~\bibnamefont
  {Rubio}},\ }\href {\doibase 10.1103/physreva.98.043801} {\bibfield  {journal}
  {\bibinfo  {journal} {Physical Review A}\ }\textbf {\bibinfo {volume} {98}}
  (\bibinfo {year} {2018}),\ 10.1103/physreva.98.043801}\BibitemShut {NoStop}%
\bibitem [{\citenamefont {Herrera}\ and\ \citenamefont
  {Spano}(2016)}]{cavity_Herrera_PRL_2016}%
  \BibitemOpen
  \bibfield  {author} {\bibinfo {author} {\bibfnamefont {F.}~\bibnamefont
  {Herrera}}\ and\ \bibinfo {author} {\bibfnamefont {F.~C.}\ \bibnamefont
  {Spano}},\ }\href {\doibase 10.1103/PhysRevLett.116.238301} {\bibfield
  {journal} {\bibinfo  {journal} {Phys. Rev. Lett.}\ }\textbf {\bibinfo
  {volume} {116}},\ \bibinfo {pages} {238301} (\bibinfo {year}
  {2016})}\BibitemShut {NoStop}%
\bibitem [{\citenamefont {Herrera}\ and\ \citenamefont
  {Spano}(2017)}]{dark_vibronic_polaritons_Herrera_PRL_2017}%
  \BibitemOpen
  \bibfield  {author} {\bibinfo {author} {\bibfnamefont {F.}~\bibnamefont
  {Herrera}}\ and\ \bibinfo {author} {\bibfnamefont {F.~C.}\ \bibnamefont
  {Spano}},\ }\href {\doibase 10.1103/PhysRevLett.118.223601} {\bibfield
  {journal} {\bibinfo  {journal} {Phys. Rev. Lett.}\ }\textbf {\bibinfo
  {volume} {118}},\ \bibinfo {pages} {223601} (\bibinfo {year}
  {2017})}\BibitemShut {NoStop}%
\bibitem [{\citenamefont {Kowalewski}\ \emph
  {et~al.}(2016{\natexlab{a}})\citenamefont {Kowalewski}, \citenamefont
  {Bennett},\ and\ \citenamefont {Mukamel}}]{Markus_1a}%
  \BibitemOpen
  \bibfield  {author} {\bibinfo {author} {\bibfnamefont {M.}~\bibnamefont
  {Kowalewski}}, \bibinfo {author} {\bibfnamefont {K.}~\bibnamefont {Bennett}},
  \ and\ \bibinfo {author} {\bibfnamefont {S.}~\bibnamefont {Mukamel}},\ }\href
  {\doibase 10.1063/1.4941053} {\bibfield  {journal} {\bibinfo  {journal} {J.
  Chem. Phys.}\ }\textbf {\bibinfo {volume} {144}},\ \bibinfo {pages} {054309}
  (\bibinfo {year} {2016}{\natexlab{a}})}\BibitemShut {NoStop}%
\bibitem [{\citenamefont {Kowalewski}\ \emph
  {et~al.}(2016{\natexlab{b}})\citenamefont {Kowalewski}, \citenamefont
  {Bennett},\ and\ \citenamefont {Mukamel}}]{cavity_Kowalewski_JPCL_2016}%
  \BibitemOpen
  \bibfield  {author} {\bibinfo {author} {\bibfnamefont {M.}~\bibnamefont
  {Kowalewski}}, \bibinfo {author} {\bibfnamefont {K.}~\bibnamefont {Bennett}},
  \ and\ \bibinfo {author} {\bibfnamefont {S.}~\bibnamefont {Mukamel}},\ }\href
  {\doibase 10.1021/acs.jpclett.6b00864} {\bibfield  {journal} {\bibinfo
  {journal} {J. Phys. Chem. Lett.}\ }\textbf {\bibinfo {volume} {7}},\ \bibinfo
  {pages} {2050} (\bibinfo {year} {2016}{\natexlab{b}})}\BibitemShut {NoStop}%
\bibitem [{\citenamefont {Davidsson}\ and\ \citenamefont
  {Kowalewski}(2020)}]{Davidsson_2020}%
  \BibitemOpen
  \bibfield  {author} {\bibinfo {author} {\bibfnamefont {E.}~\bibnamefont
  {Davidsson}}\ and\ \bibinfo {author} {\bibfnamefont {M.}~\bibnamefont
  {Kowalewski}},\ }\href {\doibase 10.1021/acs.jpca.0c03867} {\bibfield
  {journal} {\bibinfo  {journal} {The Journal of Physical Chemistry A}\
  }\textbf {\bibinfo {volume} {124}},\ \bibinfo {pages} {4672} (\bibinfo {year}
  {2020})}\BibitemShut {NoStop}%
\bibitem [{\citenamefont {Mandal}\ and\ \citenamefont
  {Huo}(2019)}]{Mandal_2019}%
  \BibitemOpen
  \bibfield  {author} {\bibinfo {author} {\bibfnamefont {A.}~\bibnamefont
  {Mandal}}\ and\ \bibinfo {author} {\bibfnamefont {P.}~\bibnamefont {Huo}},\
  }\href {\doibase 10.1021/acs.jpclett.9b01599} {\bibfield  {journal} {\bibinfo
   {journal} {The Journal of Physical Chemistry Letters}\ }\textbf {\bibinfo
  {volume} {10}},\ \bibinfo {pages} {5519} (\bibinfo {year}
  {2019})}\BibitemShut {NoStop}%
\bibitem [{\citenamefont {Ribeiro}\ \emph
  {et~al.}(2018{\natexlab{a}})\citenamefont {Ribeiro}, \citenamefont
  {Mart{\'{\i}}nez-Mart{\'{\i}}nez}, \citenamefont {Du}, \citenamefont
  {Campos-Gonzalez-Angulo},\ and\ \citenamefont {Yuen-Zhou}}]{Joel_1a}%
  \BibitemOpen
  \bibfield  {author} {\bibinfo {author} {\bibfnamefont {R.~F.}\ \bibnamefont
  {Ribeiro}}, \bibinfo {author} {\bibfnamefont {L.~A.}\ \bibnamefont
  {Mart{\'{\i}}nez-Mart{\'{\i}}nez}}, \bibinfo {author} {\bibfnamefont
  {M.}~\bibnamefont {Du}}, \bibinfo {author} {\bibfnamefont {J.}~\bibnamefont
  {Campos-Gonzalez-Angulo}}, \ and\ \bibinfo {author} {\bibfnamefont
  {J.}~\bibnamefont {Yuen-Zhou}},\ }\href {\doibase 10.1039/c8sc01043a}
  {\bibfield  {journal} {\bibinfo  {journal} {Chem. Sci.}\ }\textbf {\bibinfo
  {volume} {9}},\ \bibinfo {pages} {6325} (\bibinfo {year}
  {2018}{\natexlab{a}})}\BibitemShut {NoStop}%
\bibitem [{\citenamefont {Luk}\ \emph {et~al.}(2017)\citenamefont {Luk},
  \citenamefont {Feist}, \citenamefont {Toppari},\ and\ \citenamefont
  {Groenhof}}]{cavity_Luk_JCTC_2017}%
  \BibitemOpen
  \bibfield  {author} {\bibinfo {author} {\bibfnamefont {H.~L.}\ \bibnamefont
  {Luk}}, \bibinfo {author} {\bibfnamefont {J.}~\bibnamefont {Feist}}, \bibinfo
  {author} {\bibfnamefont {J.~J.}\ \bibnamefont {Toppari}}, \ and\ \bibinfo
  {author} {\bibfnamefont {G.}~\bibnamefont {Groenhof}},\ }\href {\doibase
  10.1021/acs.jctc.7b00388} {\bibfield  {journal} {\bibinfo  {journal} {J.
  Chem. Theory Comput.}\ }\textbf {\bibinfo {volume} {13}},\ \bibinfo {pages}
  {4324} (\bibinfo {year} {2017})}\BibitemShut {NoStop}%
\bibitem [{\citenamefont {Groenhof}\ and\ \citenamefont
  {Toppari}(2018)}]{Gerit_2a}%
  \BibitemOpen
  \bibfield  {author} {\bibinfo {author} {\bibfnamefont {G.}~\bibnamefont
  {Groenhof}}\ and\ \bibinfo {author} {\bibfnamefont {J.~J.}\ \bibnamefont
  {Toppari}},\ }\href {\doibase 10.1021/acs.jpclett.8b02032} {\bibfield
  {journal} {\bibinfo  {journal} {J. Phys. Chem. Lett.}\ }\textbf {\bibinfo
  {volume} {9}},\ \bibinfo {pages} {4848} (\bibinfo {year} {2018})}\BibitemShut
  {NoStop}%
\bibitem [{\citenamefont {Szidarovszky}\ \emph
  {et~al.}(2018{\natexlab{a}})\citenamefont {Szidarovszky}, \citenamefont
  {Hal{\'{a}}sz}, \citenamefont {Cs{\'{a}}sz{\'{a}}r}, \citenamefont
  {Cederbaum},\ and\ \citenamefont {Vib{\'{o}}k}}]{Tamas_1a}%
  \BibitemOpen
  \bibfield  {author} {\bibinfo {author} {\bibfnamefont {T.}~\bibnamefont
  {Szidarovszky}}, \bibinfo {author} {\bibfnamefont {G.~J.}\ \bibnamefont
  {Hal{\'{a}}sz}}, \bibinfo {author} {\bibfnamefont {A.~G.}\ \bibnamefont
  {Cs{\'{a}}sz{\'{a}}r}}, \bibinfo {author} {\bibfnamefont {L.~S.}\
  \bibnamefont {Cederbaum}}, \ and\ \bibinfo {author} {\bibfnamefont
  {{\'{A}}.}~\bibnamefont {Vib{\'{o}}k}},\ }\href {\doibase
  10.1021/acs.jpclett.8b02609} {\bibfield  {journal} {\bibinfo  {journal} {J.
  Phys. Chem. Lett.}\ }\textbf {\bibinfo {volume} {9}},\ \bibinfo {pages}
  {6215} (\bibinfo {year} {2018}{\natexlab{a}})}\BibitemShut {NoStop}%
\bibitem [{\citenamefont {Csehi}\ \emph
  {et~al.}(2019{\natexlab{a}})\citenamefont {Csehi}, \citenamefont
  {Kowalewski}, \citenamefont {Hal{\'{a}}sz},\ and\ \citenamefont
  {Vib{\'{o}}k}}]{Agi_2}%
  \BibitemOpen
  \bibfield  {author} {\bibinfo {author} {\bibfnamefont {A.}~\bibnamefont
  {Csehi}}, \bibinfo {author} {\bibfnamefont {M.}~\bibnamefont {Kowalewski}},
  \bibinfo {author} {\bibfnamefont {G.~J.}\ \bibnamefont {Hal{\'{a}}sz}}, \
  and\ \bibinfo {author} {\bibfnamefont {{\'{A}}.}~\bibnamefont
  {Vib{\'{o}}k}},\ }\href {\doibase 10.1088/1367-2630/ab3fcc} {\bibfield
  {journal} {\bibinfo  {journal} {New J. Phys.}\ }\textbf {\bibinfo {volume}
  {21}},\ \bibinfo {pages} {093040} (\bibinfo {year}
  {2019}{\natexlab{a}})}\BibitemShut {NoStop}%
\bibitem [{\citenamefont {Csehi}\ \emph
  {et~al.}(2019{\natexlab{b}})\citenamefont {Csehi}, \citenamefont
  {Vib{\'{o}}k}, \citenamefont {Hal{\'{a}}sz},\ and\ \citenamefont
  {Kowalewski}}]{Csehi_2019b}%
  \BibitemOpen
  \bibfield  {author} {\bibinfo {author} {\bibfnamefont {A.}~\bibnamefont
  {Csehi}}, \bibinfo {author} {\bibfnamefont {{\'{A}}.}~\bibnamefont
  {Vib{\'{o}}k}}, \bibinfo {author} {\bibfnamefont {G.~J.}\ \bibnamefont
  {Hal{\'{a}}sz}}, \ and\ \bibinfo {author} {\bibfnamefont {M.}~\bibnamefont
  {Kowalewski}},\ }\href {\doibase 10.1103/physreva.100.053421} {\bibfield
  {journal} {\bibinfo  {journal} {Physical Review A}\ }\textbf {\bibinfo
  {volume} {100}} (\bibinfo {year} {2019}{\natexlab{b}}),\
  10.1103/physreva.100.053421}\BibitemShut {NoStop}%
\bibitem [{\citenamefont
  {Vendrell}(2018{\natexlab{a}})}]{cavity_MCTDH_Vendrell_CP_2018}%
  \BibitemOpen
  \bibfield  {author} {\bibinfo {author} {\bibfnamefont {O.}~\bibnamefont
  {Vendrell}},\ }\href {\doibase
  https://doi.org/10.1016/j.chemphys.2018.02.008} {\bibfield  {journal}
  {\bibinfo  {journal} {Chem. Phys.}\ }\textbf {\bibinfo {volume} {509}},\
  \bibinfo {pages} {55 } (\bibinfo {year} {2018}{\natexlab{a}})}\BibitemShut
  {NoStop}%
\bibitem [{\citenamefont {Vendrell}(2018{\natexlab{b}})}]{Oriol_2a}%
  \BibitemOpen
  \bibfield  {author} {\bibinfo {author} {\bibfnamefont {O.}~\bibnamefont
  {Vendrell}},\ }\href {\doibase 10.1103/physrevlett.121.253001} {\bibfield
  {journal} {\bibinfo  {journal} {Phys. Rev. Lett.}\ }\textbf {\bibinfo
  {volume} {121}} (\bibinfo {year} {2018}{\natexlab{b}}),\
  10.1103/physrevlett.121.253001}\BibitemShut {NoStop}%
\bibitem [{\citenamefont {Gu}\ and\ \citenamefont
  {Mukamel}(2020{\natexlab{a}})}]{Gu_2020}%
  \BibitemOpen
  \bibfield  {author} {\bibinfo {author} {\bibfnamefont {B.}~\bibnamefont
  {Gu}}\ and\ \bibinfo {author} {\bibfnamefont {S.}~\bibnamefont {Mukamel}},\
  }\href {\doibase 10.1039/c9sc04992d} {\bibfield  {journal} {\bibinfo
  {journal} {Chemical Science}\ }\textbf {\bibinfo {volume} {11}},\ \bibinfo
  {pages} {1290} (\bibinfo {year} {2020}{\natexlab{a}})}\BibitemShut {NoStop}%
\bibitem [{\citenamefont {del Pino}\ \emph {et~al.}(2015)\citenamefont {del
  Pino}, \citenamefont {Feist},\ and\ \citenamefont
  {Garcia-Vidal}}]{Pino_2015}%
  \BibitemOpen
  \bibfield  {author} {\bibinfo {author} {\bibfnamefont {J.}~\bibnamefont {del
  Pino}}, \bibinfo {author} {\bibfnamefont {J.}~\bibnamefont {Feist}}, \ and\
  \bibinfo {author} {\bibfnamefont {F.~J.}\ \bibnamefont {Garcia-Vidal}},\
  }\href {\doibase 10.1088/1367-2630/17/5/053040} {\bibfield  {journal}
  {\bibinfo  {journal} {New Journal of Physics}\ }\textbf {\bibinfo {volume}
  {17}},\ \bibinfo {pages} {053040} (\bibinfo {year} {2015})}\BibitemShut
  {NoStop}%
\bibitem [{\citenamefont {Dunkelberger}\ \emph {et~al.}(2016)\citenamefont
  {Dunkelberger}, \citenamefont {Spann}, \citenamefont {Fears}, \citenamefont
  {Simpkins},\ and\ \citenamefont {Owrutsky}}]{Dunkelberger_2016}%
  \BibitemOpen
  \bibfield  {author} {\bibinfo {author} {\bibfnamefont {A.~D.}\ \bibnamefont
  {Dunkelberger}}, \bibinfo {author} {\bibfnamefont {B.~T.}\ \bibnamefont
  {Spann}}, \bibinfo {author} {\bibfnamefont {K.~P.}\ \bibnamefont {Fears}},
  \bibinfo {author} {\bibfnamefont {B.~S.}\ \bibnamefont {Simpkins}}, \ and\
  \bibinfo {author} {\bibfnamefont {J.~C.}\ \bibnamefont {Owrutsky}},\ }\href
  {\doibase 10.1038/ncomms13504} {\bibfield  {journal} {\bibinfo  {journal}
  {Nature Communications}\ }\textbf {\bibinfo {volume} {7}} (\bibinfo {year}
  {2016}),\ 10.1038/ncomms13504}\BibitemShut {NoStop}%
\bibitem [{\citenamefont {Herrera}\ and\ \citenamefont
  {Spano}(2018)}]{theory_of_organic_cavities_Herrera_ACSphotonics_2018}%
  \BibitemOpen
  \bibfield  {author} {\bibinfo {author} {\bibfnamefont {F.}~\bibnamefont
  {Herrera}}\ and\ \bibinfo {author} {\bibfnamefont {F.~C.}\ \bibnamefont
  {Spano}},\ }\href {\doibase 10.1021/acsphotonics.7b00728} {\bibfield
  {journal} {\bibinfo  {journal} {ACS Photonics}\ }\textbf {\bibinfo {volume}
  {5}},\ \bibinfo {pages} {65} (\bibinfo {year} {2018})}\BibitemShut {NoStop}%
\bibitem [{\citenamefont {Ahn}\ \emph {et~al.}(2018)\citenamefont {Ahn},
  \citenamefont {Vurgaftman}, \citenamefont {Dunkelberger}, \citenamefont
  {Owrutsky},\ and\ \citenamefont {Simpkins}}]{Ahn_2018}%
  \BibitemOpen
  \bibfield  {author} {\bibinfo {author} {\bibfnamefont {W.}~\bibnamefont
  {Ahn}}, \bibinfo {author} {\bibfnamefont {I.}~\bibnamefont {Vurgaftman}},
  \bibinfo {author} {\bibfnamefont {A.~D.}\ \bibnamefont {Dunkelberger}},
  \bibinfo {author} {\bibfnamefont {J.~C.}\ \bibnamefont {Owrutsky}}, \ and\
  \bibinfo {author} {\bibfnamefont {B.~S.}\ \bibnamefont {Simpkins}},\ }\href
  {\doibase 10.1021/acsphotonics.7b00583} {\bibfield  {journal} {\bibinfo
  {journal} {{ACS} Photonics}\ }\textbf {\bibinfo {volume} {5}},\ \bibinfo
  {pages} {158} (\bibinfo {year} {2018})}\BibitemShut {NoStop}%
\bibitem [{\citenamefont {Thomas}\ \emph {et~al.}(2016)\citenamefont {Thomas},
  \citenamefont {George}, \citenamefont {Shalabney}, \citenamefont {Dryzhakov},
  \citenamefont {Varma}, \citenamefont {Moran}, \citenamefont {Chervy},
  \citenamefont {Zhong}, \citenamefont {Devaux}, \citenamefont {Genet},
  \citenamefont {Hutchison},\ and\ \citenamefont {Ebbesen}}]{Ebbesen_3a}%
  \BibitemOpen
  \bibfield  {author} {\bibinfo {author} {\bibfnamefont {A.}~\bibnamefont
  {Thomas}}, \bibinfo {author} {\bibfnamefont {J.}~\bibnamefont {George}},
  \bibinfo {author} {\bibfnamefont {A.}~\bibnamefont {Shalabney}}, \bibinfo
  {author} {\bibfnamefont {M.}~\bibnamefont {Dryzhakov}}, \bibinfo {author}
  {\bibfnamefont {S.~J.}\ \bibnamefont {Varma}}, \bibinfo {author}
  {\bibfnamefont {J.}~\bibnamefont {Moran}}, \bibinfo {author} {\bibfnamefont
  {T.}~\bibnamefont {Chervy}}, \bibinfo {author} {\bibfnamefont
  {X.}~\bibnamefont {Zhong}}, \bibinfo {author} {\bibfnamefont
  {E.}~\bibnamefont {Devaux}}, \bibinfo {author} {\bibfnamefont
  {C.}~\bibnamefont {Genet}}, \bibinfo {author} {\bibfnamefont {J.~A.}\
  \bibnamefont {Hutchison}}, \ and\ \bibinfo {author} {\bibfnamefont {T.~W.}\
  \bibnamefont {Ebbesen}},\ }\href {\doibase 10.1002/anie.201605504} {\bibfield
   {journal} {\bibinfo  {journal} {Angew. Chem. Int. Ed.}\ }\textbf {\bibinfo
  {volume} {55}},\ \bibinfo {pages} {11462} (\bibinfo {year}
  {2016})}\BibitemShut {NoStop}%
\bibitem [{\citenamefont {Vergauwe}\ \emph {et~al.}(2016)\citenamefont
  {Vergauwe}, \citenamefont {George}, \citenamefont {Chervy}, \citenamefont
  {Hutchison}, \citenamefont {Shalabney}, \citenamefont {Torbeev},\ and\
  \citenamefont {Ebbesen}}]{Ebbesen_4a}%
  \BibitemOpen
  \bibfield  {author} {\bibinfo {author} {\bibfnamefont {R.~M.~A.}\
  \bibnamefont {Vergauwe}}, \bibinfo {author} {\bibfnamefont {J.}~\bibnamefont
  {George}}, \bibinfo {author} {\bibfnamefont {T.}~\bibnamefont {Chervy}},
  \bibinfo {author} {\bibfnamefont {J.~A.}\ \bibnamefont {Hutchison}}, \bibinfo
  {author} {\bibfnamefont {A.}~\bibnamefont {Shalabney}}, \bibinfo {author}
  {\bibfnamefont {V.~Y.}\ \bibnamefont {Torbeev}}, \ and\ \bibinfo {author}
  {\bibfnamefont {T.~W.}\ \bibnamefont {Ebbesen}},\ }\href {\doibase
  10.1021/acs.jpclett.6b01869} {\bibfield  {journal} {\bibinfo  {journal} {J.
  Phys. Chem. Lett.}\ }\textbf {\bibinfo {volume} {7}},\ \bibinfo {pages}
  {4159} (\bibinfo {year} {2016})}\BibitemShut {NoStop}%
\bibitem [{\citenamefont {Chervy}\ \emph {et~al.}(2018)\citenamefont {Chervy},
  \citenamefont {Thomas}, \citenamefont {Akiki}, \citenamefont {Vergauwe},
  \citenamefont {Shalabney}, \citenamefont {George}, \citenamefont {Devaux},
  \citenamefont {Hutchison}, \citenamefont {Genet},\ and\ \citenamefont
  {Ebbesen}}]{Ebbesen_5a}%
  \BibitemOpen
  \bibfield  {author} {\bibinfo {author} {\bibfnamefont {T.}~\bibnamefont
  {Chervy}}, \bibinfo {author} {\bibfnamefont {A.}~\bibnamefont {Thomas}},
  \bibinfo {author} {\bibfnamefont {E.}~\bibnamefont {Akiki}}, \bibinfo
  {author} {\bibfnamefont {R.~M.~A.}\ \bibnamefont {Vergauwe}}, \bibinfo
  {author} {\bibfnamefont {A.}~\bibnamefont {Shalabney}}, \bibinfo {author}
  {\bibfnamefont {J.}~\bibnamefont {George}}, \bibinfo {author} {\bibfnamefont
  {E.}~\bibnamefont {Devaux}}, \bibinfo {author} {\bibfnamefont {J.~A.}\
  \bibnamefont {Hutchison}}, \bibinfo {author} {\bibfnamefont {C.}~\bibnamefont
  {Genet}}, \ and\ \bibinfo {author} {\bibfnamefont {T.~W.}\ \bibnamefont
  {Ebbesen}},\ }\href {\doibase 10.1021/acsphotonics.7b00677} {\bibfield
  {journal} {\bibinfo  {journal} {{ACS} Photonics}\ }\textbf {\bibinfo {volume}
  {5}},\ \bibinfo {pages} {217} (\bibinfo {year} {2018})}\BibitemShut {NoStop}%
\bibitem [{\citenamefont {Vergauwe}\ \emph {et~al.}(2019)\citenamefont
  {Vergauwe}, \citenamefont {Thomas}, \citenamefont {Nagarajan}, \citenamefont
  {Shalabney}, \citenamefont {George}, \citenamefont {Chervy}, \citenamefont
  {Seidel}, \citenamefont {Devaux}, \citenamefont {Torbeev},\ and\
  \citenamefont {Ebbesen}}]{Ebbesen_7a}%
  \BibitemOpen
  \bibfield  {author} {\bibinfo {author} {\bibfnamefont {R.~M.~A.}\
  \bibnamefont {Vergauwe}}, \bibinfo {author} {\bibfnamefont {A.}~\bibnamefont
  {Thomas}}, \bibinfo {author} {\bibfnamefont {K.}~\bibnamefont {Nagarajan}},
  \bibinfo {author} {\bibfnamefont {A.}~\bibnamefont {Shalabney}}, \bibinfo
  {author} {\bibfnamefont {J.}~\bibnamefont {George}}, \bibinfo {author}
  {\bibfnamefont {T.}~\bibnamefont {Chervy}}, \bibinfo {author} {\bibfnamefont
  {M.}~\bibnamefont {Seidel}}, \bibinfo {author} {\bibfnamefont
  {E.}~\bibnamefont {Devaux}}, \bibinfo {author} {\bibfnamefont
  {V.}~\bibnamefont {Torbeev}}, \ and\ \bibinfo {author} {\bibfnamefont
  {T.~W.}\ \bibnamefont {Ebbesen}},\ }\href {\doibase 10.1002/anie.201908876}
  {\bibfield  {journal} {\bibinfo  {journal} {Angew. Chem. Int. Ed.}\ }\textbf
  {\bibinfo {volume} {58}},\ \bibinfo {pages} {15324} (\bibinfo {year}
  {2019})}\BibitemShut {NoStop}%
\bibitem [{\citenamefont {Thomas}\ \emph {et~al.}(2019)\citenamefont {Thomas},
  \citenamefont {Lethuillier-Karl}, \citenamefont {Nagarajan}, \citenamefont
  {Vergauwe}, \citenamefont {George}, \citenamefont {Chervy}, \citenamefont
  {Shalabney}, \citenamefont {Devaux}, \citenamefont {Genet}, \citenamefont
  {Moran},\ and\ \citenamefont {Ebbesen}}]{Thomas_2019}%
  \BibitemOpen
  \bibfield  {author} {\bibinfo {author} {\bibfnamefont {A.}~\bibnamefont
  {Thomas}}, \bibinfo {author} {\bibfnamefont {L.}~\bibnamefont
  {Lethuillier-Karl}}, \bibinfo {author} {\bibfnamefont {K.}~\bibnamefont
  {Nagarajan}}, \bibinfo {author} {\bibfnamefont {R.~M.~A.}\ \bibnamefont
  {Vergauwe}}, \bibinfo {author} {\bibfnamefont {J.}~\bibnamefont {George}},
  \bibinfo {author} {\bibfnamefont {T.}~\bibnamefont {Chervy}}, \bibinfo
  {author} {\bibfnamefont {A.}~\bibnamefont {Shalabney}}, \bibinfo {author}
  {\bibfnamefont {E.}~\bibnamefont {Devaux}}, \bibinfo {author} {\bibfnamefont
  {C.}~\bibnamefont {Genet}}, \bibinfo {author} {\bibfnamefont
  {J.}~\bibnamefont {Moran}}, \ and\ \bibinfo {author} {\bibfnamefont {T.~W.}\
  \bibnamefont {Ebbesen}},\ }\href {\doibase 10.1126/science.aau7742}
  {\bibfield  {journal} {\bibinfo  {journal} {Science}\ }\textbf {\bibinfo
  {volume} {363}},\ \bibinfo {pages} {615} (\bibinfo {year}
  {2019})}\BibitemShut {NoStop}%
\bibitem [{\citenamefont {Long}\ and\ \citenamefont
  {Simpkins}(2015)}]{Long_2014}%
  \BibitemOpen
  \bibfield  {author} {\bibinfo {author} {\bibfnamefont {J.~P.}\ \bibnamefont
  {Long}}\ and\ \bibinfo {author} {\bibfnamefont {B.~S.}\ \bibnamefont
  {Simpkins}},\ }\href {\doibase 10.1021/ph5003347} {\bibfield  {journal}
  {\bibinfo  {journal} {{ACS} Photonics}\ }\textbf {\bibinfo {volume} {2}},\
  \bibinfo {pages} {130} (\bibinfo {year} {2015})}\BibitemShut {NoStop}%
\bibitem [{\citenamefont {Muallem}\ \emph {et~al.}(2016)\citenamefont
  {Muallem}, \citenamefont {Palatnik}, \citenamefont {Nessim},\ and\
  \citenamefont {Tischler}}]{cavity_Muallem_JPCL_2016}%
  \BibitemOpen
  \bibfield  {author} {\bibinfo {author} {\bibfnamefont {M.}~\bibnamefont
  {Muallem}}, \bibinfo {author} {\bibfnamefont {A.}~\bibnamefont {Palatnik}},
  \bibinfo {author} {\bibfnamefont {G.~D.}\ \bibnamefont {Nessim}}, \ and\
  \bibinfo {author} {\bibfnamefont {Y.~R.}\ \bibnamefont {Tischler}},\ }\href
  {\doibase 10.1021/acs.jpclett.6b00617} {\bibfield  {journal} {\bibinfo
  {journal} {J. Phys. Chem. Lett.}\ }\textbf {\bibinfo {volume} {7}},\ \bibinfo
  {pages} {2002} (\bibinfo {year} {2016})}\BibitemShut {NoStop}%
\bibitem [{\citenamefont {Damari}\ \emph {et~al.}(2019)\citenamefont {Damari},
  \citenamefont {Weinberg}, \citenamefont {Krotkov}, \citenamefont {Demina},
  \citenamefont {Akulov}, \citenamefont {Golombek}, \citenamefont {Schwartz},\
  and\ \citenamefont {Fleischer}}]{Fleischer_2019}%
  \BibitemOpen
  \bibfield  {author} {\bibinfo {author} {\bibfnamefont {R.}~\bibnamefont
  {Damari}}, \bibinfo {author} {\bibfnamefont {O.}~\bibnamefont {Weinberg}},
  \bibinfo {author} {\bibfnamefont {D.}~\bibnamefont {Krotkov}}, \bibinfo
  {author} {\bibfnamefont {N.}~\bibnamefont {Demina}}, \bibinfo {author}
  {\bibfnamefont {K.}~\bibnamefont {Akulov}}, \bibinfo {author} {\bibfnamefont
  {A.}~\bibnamefont {Golombek}}, \bibinfo {author} {\bibfnamefont
  {T.}~\bibnamefont {Schwartz}}, \ and\ \bibinfo {author} {\bibfnamefont
  {S.}~\bibnamefont {Fleischer}},\ }\href {\doibase 10.1038/s41467-019-11130-y}
  {\bibfield  {journal} {\bibinfo  {journal} {Nature Communications}\ }\textbf
  {\bibinfo {volume} {10}},\ \bibinfo {pages} {3248} (\bibinfo {year}
  {2019})}\BibitemShut {NoStop}%
\bibitem [{\citenamefont {Campos-Gonzalez-Angulo}\ \emph
  {et~al.}(2019)\citenamefont {Campos-Gonzalez-Angulo}, \citenamefont
  {Ribeiro},\ and\ \citenamefont {Yuen-Zhou}}]{Joel_4a}%
  \BibitemOpen
  \bibfield  {author} {\bibinfo {author} {\bibfnamefont {J.~A.}\ \bibnamefont
  {Campos-Gonzalez-Angulo}}, \bibinfo {author} {\bibfnamefont {R.~F.}\
  \bibnamefont {Ribeiro}}, \ and\ \bibinfo {author} {\bibfnamefont
  {J.}~\bibnamefont {Yuen-Zhou}},\ }\href {\doibase 10.1038/s41467-019-12636-1}
  {\bibfield  {journal} {\bibinfo  {journal} {Nat. Commun.}\ }\textbf {\bibinfo
  {volume} {10}},\ \bibinfo {pages} {1} (\bibinfo {year} {2019})}\BibitemShut
  {NoStop}%
\bibitem [{\citenamefont {K{\'{e}}na-Cohen}\ and\ \citenamefont
  {Yuen-Zhou}(2019)}]{Joel_3a}%
  \BibitemOpen
  \bibfield  {author} {\bibinfo {author} {\bibfnamefont {S.}~\bibnamefont
  {K{\'{e}}na-Cohen}}\ and\ \bibinfo {author} {\bibfnamefont {J.}~\bibnamefont
  {Yuen-Zhou}},\ }\href {\doibase 10.1021/acscentsci.9b00219} {\bibfield
  {journal} {\bibinfo  {journal} {{ACS} Central Science}\ }\textbf {\bibinfo
  {volume} {5}},\ \bibinfo {pages} {386} (\bibinfo {year} {2019})}\BibitemShut
  {NoStop}%
\bibitem [{\citenamefont {Mandal}\ \emph {et~al.}(2020)\citenamefont {Mandal},
  \citenamefont {Vega},\ and\ \citenamefont {Huo}}]{Mandal2020}%
  \BibitemOpen
  \bibfield  {author} {\bibinfo {author} {\bibfnamefont {A.}~\bibnamefont
  {Mandal}}, \bibinfo {author} {\bibfnamefont {S.~M.}\ \bibnamefont {Vega}}, \
  and\ \bibinfo {author} {\bibfnamefont {P.}~\bibnamefont {Huo}},\ }\href
  {\doibase 10.1021/acs.jpclett.0c02399} {\bibfield  {journal} {\bibinfo
  {journal} {The Journal of Physical Chemistry Letters}\ } (\bibinfo {year}
  {2020}),\ 10.1021/acs.jpclett.0c02399}\BibitemShut {NoStop}%
\bibitem [{\citenamefont {Ulusoy}\ \emph {et~al.}(2019)\citenamefont {Ulusoy},
  \citenamefont {Gomez},\ and\ \citenamefont {Vendrell}}]{Ulusoy2019}%
  \BibitemOpen
  \bibfield  {author} {\bibinfo {author} {\bibfnamefont {I.~S.}\ \bibnamefont
  {Ulusoy}}, \bibinfo {author} {\bibfnamefont {J.~A.}\ \bibnamefont {Gomez}}, \
  and\ \bibinfo {author} {\bibfnamefont {O.}~\bibnamefont {Vendrell}},\ }\href
  {\doibase 10.1021/acs.jpca.9b07404} {\bibfield  {journal} {\bibinfo
  {journal} {The Journal of Physical Chemistry A}\ }\textbf {\bibinfo {volume}
  {123}},\ \bibinfo {pages} {8832} (\bibinfo {year} {2019})}\BibitemShut
  {NoStop}%
\bibitem [{\citenamefont {Du}\ \emph {et~al.}(2019)\citenamefont {Du},
  \citenamefont {Ribeiro},\ and\ \citenamefont {Yuen-Zhou}}]{Du2019}%
  \BibitemOpen
  \bibfield  {author} {\bibinfo {author} {\bibfnamefont {M.}~\bibnamefont
  {Du}}, \bibinfo {author} {\bibfnamefont {R.~F.}\ \bibnamefont {Ribeiro}}, \
  and\ \bibinfo {author} {\bibfnamefont {J.}~\bibnamefont {Yuen-Zhou}},\ }\href
  {\doibase 10.1016/j.chempr.2019.02.009} {\bibfield  {journal} {\bibinfo
  {journal} {Chem}\ }\textbf {\bibinfo {volume} {5}},\ \bibinfo {pages} {1167}
  (\bibinfo {year} {2019})}\BibitemShut {NoStop}%
\bibitem [{\citenamefont {Gu}\ and\ \citenamefont
  {Mukamel}(2020{\natexlab{b}})}]{Gu2020_JPCL}%
  \BibitemOpen
  \bibfield  {author} {\bibinfo {author} {\bibfnamefont {B.}~\bibnamefont
  {Gu}}\ and\ \bibinfo {author} {\bibfnamefont {S.}~\bibnamefont {Mukamel}},\
  }\href {\doibase 10.1021/acs.jpclett.0c00381} {\bibfield  {journal} {\bibinfo
   {journal} {The Journal of Physical Chemistry Letters}\ }\textbf {\bibinfo
  {volume} {11}},\ \bibinfo {pages} {5555} (\bibinfo {year}
  {2020}{\natexlab{b}})}\BibitemShut {NoStop}%
\bibitem [{\citenamefont {Szidarovszky}\ \emph {et~al.}(2020)\citenamefont
  {Szidarovszky}, \citenamefont {Hal{\'{a}}sz},\ and\ \citenamefont
  {Vib{\'{o}}k}}]{Szidarovszky2020}%
  \BibitemOpen
  \bibfield  {author} {\bibinfo {author} {\bibfnamefont {T.}~\bibnamefont
  {Szidarovszky}}, \bibinfo {author} {\bibfnamefont {G.~J.}\ \bibnamefont
  {Hal{\'{a}}sz}}, \ and\ \bibinfo {author} {\bibfnamefont
  {{\'{A}}.}~\bibnamefont {Vib{\'{o}}k}},\ }\href {\doibase
  10.1088/1367-2630/ab8264} {\bibfield  {journal} {\bibinfo  {journal} {New
  Journal of Physics}\ }\textbf {\bibinfo {volume} {22}},\ \bibinfo {pages}
  {053001} (\bibinfo {year} {2020})}\BibitemShut {NoStop}%
\bibitem [{\citenamefont {Ribeiro}\ and\ \citenamefont
  {Yuen-Zhou}(2017)}]{Ribeiro2017}%
  \BibitemOpen
  \bibfield  {author} {\bibinfo {author} {\bibfnamefont {R.~F.}\ \bibnamefont
  {Ribeiro}}\ and\ \bibinfo {author} {\bibfnamefont {J.}~\bibnamefont
  {Yuen-Zhou}},\ }\href {\doibase 10.1021/acs.jpclett.7b02592} {\bibfield
  {journal} {\bibinfo  {journal} {The Journal of Physical Chemistry Letters}\
  }\textbf {\bibinfo {volume} {9}},\ \bibinfo {pages} {242} (\bibinfo {year}
  {2017})}\BibitemShut {NoStop}%
\bibitem [{\citenamefont {Hern{\'{a}}ndez}\ and\ \citenamefont
  {Herrera}(2019)}]{Hernandez2019}%
  \BibitemOpen
  \bibfield  {author} {\bibinfo {author} {\bibfnamefont {F.~J.}\ \bibnamefont
  {Hern{\'{a}}ndez}}\ and\ \bibinfo {author} {\bibfnamefont {F.}~\bibnamefont
  {Herrera}},\ }\href {\doibase 10.1063/1.5121426} {\bibfield  {journal}
  {\bibinfo  {journal} {The Journal of Chemical Physics}\ }\textbf {\bibinfo
  {volume} {151}},\ \bibinfo {pages} {144116} (\bibinfo {year}
  {2019})}\BibitemShut {NoStop}%
\bibitem [{\citenamefont {Triana}\ \emph {et~al.}(2020)\citenamefont {Triana},
  \citenamefont {Hern{\'{a}}ndez},\ and\ \citenamefont {Herrera}}]{Triana2020}%
  \BibitemOpen
  \bibfield  {author} {\bibinfo {author} {\bibfnamefont {J.~F.}\ \bibnamefont
  {Triana}}, \bibinfo {author} {\bibfnamefont {F.~J.}\ \bibnamefont
  {Hern{\'{a}}ndez}}, \ and\ \bibinfo {author} {\bibfnamefont {F.}~\bibnamefont
  {Herrera}},\ }\href {\doibase 10.1063/5.0009869} {\bibfield  {journal}
  {\bibinfo  {journal} {The Journal of Chemical Physics}\ }\textbf {\bibinfo
  {volume} {152}},\ \bibinfo {pages} {234111} (\bibinfo {year}
  {2020})}\BibitemShut {NoStop}%
\bibitem [{\citenamefont {K\"oppel}\ \emph {et~al.}(1984)\citenamefont
  {K\"oppel}, \citenamefont {Domcke},\ and\ \citenamefont
  {Cederbaum}}]{Cederbaum_multimode}%
  \BibitemOpen
  \bibfield  {author} {\bibinfo {author} {\bibfnamefont {H.}~\bibnamefont
  {K\"oppel}}, \bibinfo {author} {\bibfnamefont {W.}~\bibnamefont {Domcke}}, \
  and\ \bibinfo {author} {\bibfnamefont {L.~S.}\ \bibnamefont {Cederbaum}},\
  }\href@noop {} {\bibfield  {journal} {\bibinfo  {journal} {Adv. Chem. Phys.}\
  }\textbf {\bibinfo {volume} {57}},\ \bibinfo {pages} {59} (\bibinfo {year}
  {1984})}\BibitemShut {NoStop}%
\bibitem [{\citenamefont {Yarkony}(1996)}]{Yarkony_1996}%
  \BibitemOpen
  \bibfield  {author} {\bibinfo {author} {\bibfnamefont {D.~R.}\ \bibnamefont
  {Yarkony}},\ }\href {\doibase 10.1103/RevModPhys.68.985} {\bibfield
  {journal} {\bibinfo  {journal} {Rev. Mod. Phys.}\ }\textbf {\bibinfo {volume}
  {68}},\ \bibinfo {pages} {985} (\bibinfo {year} {1996})}\BibitemShut
  {NoStop}%
\bibitem [{\citenamefont {Baer}(2002)}]{Baer_2002}%
  \BibitemOpen
  \bibfield  {author} {\bibinfo {author} {\bibfnamefont {M.}~\bibnamefont
  {Baer}},\ }\href {\doibase 10.1016/s0370-1573(01)00052-7} {\bibfield
  {journal} {\bibinfo  {journal} {Physics Reports}\ }\textbf {\bibinfo {volume}
  {358}},\ \bibinfo {pages} {75} (\bibinfo {year} {2002})}\BibitemShut
  {NoStop}%
\bibitem [{\citenamefont {Worth}\ and\ \citenamefont
  {Cederbaum}(2004)}]{vibronic_coupling_model_Cederbaum_AnnRevPhysChem_2004}%
  \BibitemOpen
  \bibfield  {author} {\bibinfo {author} {\bibfnamefont {G.~A.}\ \bibnamefont
  {Worth}}\ and\ \bibinfo {author} {\bibfnamefont {L.~S.}\ \bibnamefont
  {Cederbaum}},\ }\href {\doibase 10.1146/annurev.physchem.55.091602.094335}
  {\bibfield  {journal} {\bibinfo  {journal} {Ann. Rev. Phys. Chem.}\ }\textbf
  {\bibinfo {volume} {55}},\ \bibinfo {pages} {127} (\bibinfo {year}
  {2004})}\BibitemShut {NoStop}%
\bibitem [{\citenamefont {Domcke}\ \emph {et~al.}(2004)\citenamefont {Domcke},
  \citenamefont {Yarkony},\ and\ \citenamefont {K\"{o}ppel}}]{Domcke_2004}%
  \BibitemOpen
  \bibfield  {author} {\bibinfo {author} {\bibfnamefont {W.}~\bibnamefont
  {Domcke}}, \bibinfo {author} {\bibfnamefont {D.~R.}\ \bibnamefont {Yarkony}},
  \ and\ \bibinfo {author} {\bibfnamefont {H.}~\bibnamefont {K\"{o}ppel}},\
  }\href {\doibase 10.1142/5406} {\emph {\bibinfo {title} {Conical
  Intersections: Electronic Structure, Dynamics and Spectroscopy}}}\ (\bibinfo
  {publisher} {World Scientific, Singapore},\ \bibinfo {year}
  {2004})\BibitemShut {NoStop}%
\bibitem [{\citenamefont {Baer}(2006)}]{Baer_2006}%
  \BibitemOpen
  \bibfield  {author} {\bibinfo {author} {\bibfnamefont {M.}~\bibnamefont
  {Baer}},\ }\href {\doibase 10.1002/0471780081} {\emph {\bibinfo {title}
  {Beyond Born-Oppenheimer: Electronic Non-Adiabatic Coupling Terms and Conical
  Intersections}}}\ (\bibinfo  {publisher} {John Wiley {\&} Sons, Inc., New
  York},\ \bibinfo {year} {2006})\BibitemShut {NoStop}%
\bibitem [{\citenamefont {Moiseyev}\ \emph {et~al.}(2008)\citenamefont
  {Moiseyev}, \citenamefont {\v{S}indelka},\ and\ \citenamefont
  {Cederbaum}}]{Nimrod_1}%
  \BibitemOpen
  \bibfield  {author} {\bibinfo {author} {\bibfnamefont {N.}~\bibnamefont
  {Moiseyev}}, \bibinfo {author} {\bibfnamefont {M.}~\bibnamefont
  {\v{S}indelka}}, \ and\ \bibinfo {author} {\bibfnamefont {L.~S.}\
  \bibnamefont {Cederbaum}},\ }\href
  {http://stacks.iop.org/0953-4075/41/i=22/a=221001} {\bibfield  {journal}
  {\bibinfo  {journal} {J. Phys. B}\ }\textbf {\bibinfo {volume} {41}},\
  \bibinfo {pages} {221001} (\bibinfo {year} {2008})}\BibitemShut {NoStop}%
\bibitem [{\citenamefont {Hal\'asz}\ \emph {et~al.}(2011)\citenamefont
  {Hal\'asz}, \citenamefont {Vib\'ok}, \citenamefont {\v{S}indelka},
  \citenamefont {Moiseyev},\ and\ \citenamefont {Cederbaum}}]{LICI1}%
  \BibitemOpen
  \bibfield  {author} {\bibinfo {author} {\bibfnamefont {G.~J.}\ \bibnamefont
  {Hal\'asz}}, \bibinfo {author} {\bibfnamefont {A.}~\bibnamefont {Vib\'ok}},
  \bibinfo {author} {\bibfnamefont {M.}~\bibnamefont {\v{S}indelka}}, \bibinfo
  {author} {\bibfnamefont {N.}~\bibnamefont {Moiseyev}}, \ and\ \bibinfo
  {author} {\bibfnamefont {L.~S.}\ \bibnamefont {Cederbaum}},\ }\href
  {http://stacks.iop.org/0953-4075/44/i=17/a=175102} {\bibfield  {journal}
  {\bibinfo  {journal} {J. Phys. B}\ }\textbf {\bibinfo {volume} {44}},\
  \bibinfo {pages} {175102} (\bibinfo {year} {2011})}\BibitemShut {NoStop}%
\bibitem [{\citenamefont {Hal\'asz}\ \emph {et~al.}(2012)\citenamefont
  {Hal\'asz}, \citenamefont {\v{S}indelka}, \citenamefont {Moiseyev},
  \citenamefont {Cederbaum},\ and\ \citenamefont {Vib\'ok}}]{LICI2}%
  \BibitemOpen
  \bibfield  {author} {\bibinfo {author} {\bibfnamefont {G.~J.}\ \bibnamefont
  {Hal\'asz}}, \bibinfo {author} {\bibfnamefont {M.}~\bibnamefont
  {\v{S}indelka}}, \bibinfo {author} {\bibfnamefont {N.}~\bibnamefont
  {Moiseyev}}, \bibinfo {author} {\bibfnamefont {L.~S.}\ \bibnamefont
  {Cederbaum}}, \ and\ \bibinfo {author} {\bibfnamefont {A.}~\bibnamefont
  {Vib\'ok}},\ }\href {\doibase 10.1021/jp206860p} {\bibfield  {journal}
  {\bibinfo  {journal} {J. Phys. Chem. A}\ }\textbf {\bibinfo {volume} {116}},\
  \bibinfo {pages} {2636} (\bibinfo {year} {2012})}\BibitemShut {NoStop}%
\bibitem [{\citenamefont {Hal\'asz}\ \emph {et~al.}(2014)\citenamefont
  {Hal\'asz}, \citenamefont {Csehi}, \citenamefont {Vib\'ok},\ and\
  \citenamefont {Cederbaum}}]{LICI3}%
  \BibitemOpen
  \bibfield  {author} {\bibinfo {author} {\bibfnamefont {G.~J.}\ \bibnamefont
  {Hal\'asz}}, \bibinfo {author} {\bibfnamefont {A.}~\bibnamefont {Csehi}},
  \bibinfo {author} {\bibfnamefont {A.}~\bibnamefont {Vib\'ok}}, \ and\
  \bibinfo {author} {\bibfnamefont {L.~S.}\ \bibnamefont {Cederbaum}},\ }\href
  {\doibase 10.1021/jp504889e} {\bibfield  {journal} {\bibinfo  {journal} {J.
  Phys. Chem. A}\ }\textbf {\bibinfo {volume} {118}},\ \bibinfo {pages} {11908}
  (\bibinfo {year} {2014})}\BibitemShut {NoStop}%
\bibitem [{\citenamefont {Hal\'asz}\ \emph {et~al.}(2015)\citenamefont
  {Hal\'asz}, \citenamefont {Vib\'ok},\ and\ \citenamefont
  {Cederbaum}}]{LICI5}%
  \BibitemOpen
  \bibfield  {author} {\bibinfo {author} {\bibfnamefont {G.~J.}\ \bibnamefont
  {Hal\'asz}}, \bibinfo {author} {\bibfnamefont {A.}~\bibnamefont {Vib\'ok}}, \
  and\ \bibinfo {author} {\bibfnamefont {L.~S.}\ \bibnamefont {Cederbaum}},\
  }\href {\doibase 10.1021/jz502468d} {\bibfield  {journal} {\bibinfo
  {journal} {J. Phys. Chem. Lett.}\ }\textbf {\bibinfo {volume} {6}},\ \bibinfo
  {pages} {348} (\bibinfo {year} {2015})}\BibitemShut {NoStop}%
\bibitem [{\citenamefont {Szidarovszky}\ \emph
  {et~al.}(2018{\natexlab{b}})\citenamefont {Szidarovszky}, \citenamefont
  {Hal{\'{a}}sz}, \citenamefont {Cs{\'{a}}sz{\'{a}}r}, \citenamefont
  {Cederbaum},\ and\ \citenamefont
  {Vib{\'{o}}k}}]{LICI_in_spectrum_Szidarovszky_JPCL_2018}%
  \BibitemOpen
  \bibfield  {author} {\bibinfo {author} {\bibfnamefont {T.}~\bibnamefont
  {Szidarovszky}}, \bibinfo {author} {\bibfnamefont {G.~J.}\ \bibnamefont
  {Hal{\'{a}}sz}}, \bibinfo {author} {\bibfnamefont {A.~G.}\ \bibnamefont
  {Cs{\'{a}}sz{\'{a}}r}}, \bibinfo {author} {\bibfnamefont {L.~S.}\
  \bibnamefont {Cederbaum}}, \ and\ \bibinfo {author} {\bibfnamefont
  {{\'{A}}.}~\bibnamefont {Vib{\'{o}}k}},\ }\href {\doibase
  10.1021/acs.jpclett.8b01102} {\bibfield  {journal} {\bibinfo  {journal} {J.
  Phys. Chem. Lett.}\ }\textbf {\bibinfo {volume} {9}},\ \bibinfo {pages}
  {2739} (\bibinfo {year} {2018}{\natexlab{b}})}\BibitemShut {NoStop}%
\bibitem [{\citenamefont {Szidarovszky}\ \emph {et~al.}(2010)\citenamefont
  {Szidarovszky}, \citenamefont {Cs{\'{a}}sz{\'{a}}r},\ and\ \citenamefont
  {Czak{\'{o}}}}]{D2FOPI_Szidarovszky_PCCP_2010}%
  \BibitemOpen
  \bibfield  {author} {\bibinfo {author} {\bibfnamefont {T.}~\bibnamefont
  {Szidarovszky}}, \bibinfo {author} {\bibfnamefont {A.~G.}\ \bibnamefont
  {Cs{\'{a}}sz{\'{a}}r}}, \ and\ \bibinfo {author} {\bibfnamefont
  {G.}~\bibnamefont {Czak{\'{o}}}},\ }\href {\doibase 10.1039/c001124j}
  {\bibfield  {journal} {\bibinfo  {journal} {Phys. Chem. Chem. Phys.}\
  }\textbf {\bibinfo {volume} {12}},\ \bibinfo {pages} {8373} (\bibinfo {year}
  {2010})}\BibitemShut {NoStop}%
\bibitem [{\citenamefont {Magnier}\ \emph {et~al.}(1993)\citenamefont
  {Magnier}, \citenamefont {Milli\'e}, \citenamefont {Dulieu},\ and\
  \citenamefont {Masnou-Seeuws}}]{Na2_PEC_Magnier_JCP_1993}%
  \BibitemOpen
  \bibfield  {author} {\bibinfo {author} {\bibfnamefont {S.}~\bibnamefont
  {Magnier}}, \bibinfo {author} {\bibfnamefont {P.}~\bibnamefont {Milli\'e}},
  \bibinfo {author} {\bibfnamefont {O.}~\bibnamefont {Dulieu}}, \ and\ \bibinfo
  {author} {\bibfnamefont {F.}~\bibnamefont {Masnou-Seeuws}},\ }\href@noop {}
  {\bibfield  {journal} {\bibinfo  {journal} {J. Chem. Phys.}\ }\textbf
  {\bibinfo {volume} {98}},\ \bibinfo {pages} {7113} (\bibinfo {year}
  {1993})}\BibitemShut {NoStop}%
\bibitem [{\citenamefont {Zemke}\ \emph {et~al.}(1981)\citenamefont {Zemke},
  \citenamefont {Verma}, \citenamefont {Vu},\ and\ \citenamefont
  {Stwalley}}]{Na2_TDM_Zemke_JMS_1981}%
  \BibitemOpen
  \bibfield  {author} {\bibinfo {author} {\bibfnamefont {W.~T.}\ \bibnamefont
  {Zemke}}, \bibinfo {author} {\bibfnamefont {K.~K.}\ \bibnamefont {Verma}},
  \bibinfo {author} {\bibfnamefont {T.}~\bibnamefont {Vu}}, \ and\ \bibinfo
  {author} {\bibfnamefont {W.~C.}\ \bibnamefont {Stwalley}},\ }\href@noop {}
  {\bibfield  {journal} {\bibinfo  {journal} {J. Mol. Spectrosc.}\ }\textbf
  {\bibinfo {volume} {85}},\ \bibinfo {pages} {150} (\bibinfo {year}
  {1981})}\BibitemShut {NoStop}%
\bibitem [{\citenamefont {Zare}(1988)}]{88Zare}%
  \BibitemOpen
  \bibfield  {author} {\bibinfo {author} {\bibfnamefont {R.~N.}\ \bibnamefont
  {Zare}},\ }\href@noop {} {\emph {\bibinfo {title} {Angular Momentum:
  Understanding Spatial Aspects in Chemistry and Physics}}}\ (\bibinfo
  {publisher} {Wiley--Interscience},\ \bibinfo {address} {New York},\ \bibinfo
  {year} {1988})\BibitemShut {NoStop}%
\bibitem [{\citenamefont {Houdr{\'{e}}}\ \emph {et~al.}(1996)\citenamefont
  {Houdr{\'{e}}}, \citenamefont {Stanley},\ and\ \citenamefont
  {Ilegems}}]{Houdr1996}%
  \BibitemOpen
  \bibfield  {author} {\bibinfo {author} {\bibfnamefont {R.}~\bibnamefont
  {Houdr{\'{e}}}}, \bibinfo {author} {\bibfnamefont {R.~P.}\ \bibnamefont
  {Stanley}}, \ and\ \bibinfo {author} {\bibfnamefont {M.}~\bibnamefont
  {Ilegems}},\ }\href {\doibase 10.1103/physreva.53.2711} {\bibfield  {journal}
  {\bibinfo  {journal} {Phys. Rev. A}\ }\textbf {\bibinfo {volume} {53}},\
  \bibinfo {pages} {2711} (\bibinfo {year} {1996})}\BibitemShut {NoStop}%
\bibitem [{\citenamefont {Longuet-Higgins}\ \emph {et~al.}(1958)\citenamefont
  {Longuet-Higgins}, \citenamefont {Opik}, \citenamefont {Pryce},\ and\
  \citenamefont {Sack}}]{Longuet-Higgins1958}%
  \BibitemOpen
  \bibfield  {author} {\bibinfo {author} {\bibfnamefont {H.~C.}\ \bibnamefont
  {Longuet-Higgins}}, \bibinfo {author} {\bibfnamefont {U.}~\bibnamefont
  {Opik}}, \bibinfo {author} {\bibfnamefont {M.~H.~L.}\ \bibnamefont {Pryce}},
  \ and\ \bibinfo {author} {\bibfnamefont {R.~A.}\ \bibnamefont {Sack}},\
  }\href {\doibase 10.1098/rspa.1958.0022} {\bibfield  {journal} {\bibinfo
  {journal} {Proceedings of the Royal Society of London. Series A. Mathematical
  and Physical Sciences}\ }\textbf {\bibinfo {volume} {244}},\ \bibinfo {pages}
  {1} (\bibinfo {year} {1958})}\BibitemShut {NoStop}%
\bibitem [{\citenamefont {Longuet-Higgins}(1961)}]{Longuet-Higgins1961429}%
  \BibitemOpen
  \bibfield  {author} {\bibinfo {author} {\bibfnamefont {H.}~\bibnamefont
  {Longuet-Higgins}},\ }\href@noop {} {\bibfield  {journal} {\bibinfo
  {journal} {Adv. Spectrosc. II}\ ,\ \bibinfo {pages} {429}} (\bibinfo {year}
  {1961})},\ \bibinfo {note} {cited By 1}\BibitemShut {NoStop}%
\bibitem [{\citenamefont {Mead}\ and\ \citenamefont
  {Truhlar}(1982)}]{Mead1982}%
  \BibitemOpen
  \bibfield  {author} {\bibinfo {author} {\bibfnamefont {C.~A.}\ \bibnamefont
  {Mead}}\ and\ \bibinfo {author} {\bibfnamefont {D.~G.}\ \bibnamefont
  {Truhlar}},\ }\href {\doibase 10.1063/1.443853} {\bibfield  {journal}
  {\bibinfo  {journal} {The Journal of Chemical Physics}\ }\textbf {\bibinfo
  {volume} {77}},\ \bibinfo {pages} {6090} (\bibinfo {year}
  {1982})}\BibitemShut {NoStop}%
\bibitem [{\citenamefont {Berry}(1984)}]{Berry1984}%
  \BibitemOpen
  \bibfield  {author} {\bibinfo {author} {\bibfnamefont {M.~V.}\ \bibnamefont
  {Berry}},\ }\href {\doibase 10.1098/rspa.1984.0023} {\bibfield  {journal}
  {\bibinfo  {journal} {Proceedings of the Royal Society of London. A.
  Mathematical and Physical Sciences}\ }\textbf {\bibinfo {volume} {392}},\
  \bibinfo {pages} {45} (\bibinfo {year} {1984})}\BibitemShut {NoStop}%
\bibitem [{\citenamefont {Hal{\'{a}}sz}\ \emph {et~al.}(2018)\citenamefont
  {Hal{\'{a}}sz}, \citenamefont {Badank{\'{o}}},\ and\ \citenamefont
  {Vib{\'{o}}k}}]{Halsz2018}%
  \BibitemOpen
  \bibfield  {author} {\bibinfo {author} {\bibfnamefont {G.~J.}\ \bibnamefont
  {Hal{\'{a}}sz}}, \bibinfo {author} {\bibfnamefont {P.}~\bibnamefont
  {Badank{\'{o}}}}, \ and\ \bibinfo {author} {\bibfnamefont
  {{\'{A}}.}~\bibnamefont {Vib{\'{o}}k}},\ }\href {\doibase
  10.1080/00268976.2018.1431410} {\bibfield  {journal} {\bibinfo  {journal}
  {Molecular Physics}\ }\textbf {\bibinfo {volume} {116}},\ \bibinfo {pages}
  {2652} (\bibinfo {year} {2018})}\BibitemShut {NoStop}%
\bibitem [{\citenamefont {Baer}(1975)}]{Baer1975112}%
  \BibitemOpen
  \bibfield  {author} {\bibinfo {author} {\bibfnamefont {M.}~\bibnamefont
  {Baer}},\ }\href
  {http://www.scopus.com/inward/record.url?eid=2-s2.0-0000677123&partnerID=40&md5=4b62e262324397dbecda0269250d4102}
  {\bibfield  {journal} {\bibinfo  {journal} {Chemical Physics Letters}\
  }\textbf {\bibinfo {volume} {35}},\ \bibinfo {pages} {112} (\bibinfo {year}
  {1975})},\ \bibinfo {note} {cited By (since 1996)225}\BibitemShut {NoStop}%
\bibitem [{\citenamefont {V\'ertesi}\ \emph {et~al.}(2005)\citenamefont
  {V\'ertesi}, \citenamefont {Bene}, \citenamefont {Vib\'ok}, \citenamefont
  {Hal\'asz},\ and\ \citenamefont {Baer}}]{Vertesi20053476}%
  \BibitemOpen
  \bibfield  {author} {\bibinfo {author} {\bibfnamefont {T.}~\bibnamefont
  {V\'ertesi}}, \bibinfo {author} {\bibfnamefont {E.}~\bibnamefont {Bene}},
  \bibinfo {author} {\bibfnamefont {A.}~\bibnamefont {Vib\'ok}}, \bibinfo
  {author} {\bibfnamefont {G.~J.}\ \bibnamefont {Hal\'asz}}, \ and\ \bibinfo
  {author} {\bibfnamefont {M.}~\bibnamefont {Baer}},\ }\href
  {http://www.scopus.com/inward/record.url?eid=2-s2.0-18144428352&partnerID=40&md5=8f07ae14753a371cdf7e5a00203bbb2d}
  {\bibfield  {journal} {\bibinfo  {journal} {Journal of Physical Chemistry A}\
  }\textbf {\bibinfo {volume} {109}},\ \bibinfo {pages} {3476} (\bibinfo {year}
  {2005})},\ \bibinfo {note} {cited By (since 1996)19}\BibitemShut {NoStop}%
\bibitem [{\citenamefont {Zeb}\ \emph {et~al.}(2018)\citenamefont {Zeb},
  \citenamefont {Kirton},\ and\ \citenamefont
  {Keeling}}]{spectra_of_vibDressedpolaritons_Zeb_ACSphotonics_2018}%
  \BibitemOpen
  \bibfield  {author} {\bibinfo {author} {\bibfnamefont {M.~A.}\ \bibnamefont
  {Zeb}}, \bibinfo {author} {\bibfnamefont {P.~G.}\ \bibnamefont {Kirton}}, \
  and\ \bibinfo {author} {\bibfnamefont {J.}~\bibnamefont {Keeling}},\ }\href
  {\doibase 10.1021/acsphotonics.7b00916} {\bibfield  {journal} {\bibinfo
  {journal} {ACS Photonics}\ }\textbf {\bibinfo {volume} {5}},\ \bibinfo
  {pages} {249} (\bibinfo {year} {2018})}\BibitemShut {NoStop}%
\bibitem [{\citenamefont {Ribeiro}\ \emph
  {et~al.}(2018{\natexlab{b}})\citenamefont {Ribeiro}, \citenamefont
  {Dunkelberger}, \citenamefont {Xiang}, \citenamefont {Xiong}, \citenamefont
  {Simpkins}, \citenamefont {Owrutsky},\ and\ \citenamefont
  {Yuen-Zhou}}]{FRibeiro2018}%
  \BibitemOpen
  \bibfield  {author} {\bibinfo {author} {\bibfnamefont {R.~F.}\ \bibnamefont
  {Ribeiro}}, \bibinfo {author} {\bibfnamefont {A.~D.}\ \bibnamefont
  {Dunkelberger}}, \bibinfo {author} {\bibfnamefont {B.}~\bibnamefont {Xiang}},
  \bibinfo {author} {\bibfnamefont {W.}~\bibnamefont {Xiong}}, \bibinfo
  {author} {\bibfnamefont {B.~S.}\ \bibnamefont {Simpkins}}, \bibinfo {author}
  {\bibfnamefont {J.~C.}\ \bibnamefont {Owrutsky}}, \ and\ \bibinfo {author}
  {\bibfnamefont {J.}~\bibnamefont {Yuen-Zhou}},\ }\href {\doibase
  10.1021/acs.jpclett.8b01176} {\bibfield  {journal} {\bibinfo  {journal} {The
  Journal of Physical Chemistry Letters}\ }\textbf {\bibinfo {volume} {9}},\
  \bibinfo {pages} {3766} (\bibinfo {year} {2018}{\natexlab{b}})}\BibitemShut
  {NoStop}%
\bibitem [{\citenamefont {Szidarovszky}\ \emph {et~al.}(2019)\citenamefont
  {Szidarovszky}, \citenamefont {Cs\'asz\'ar}, \citenamefont {Hal\'asz},\ and\
  \citenamefont {Vib\'ok}}]{Szidarovszky2019}%
  \BibitemOpen
  \bibfield  {author} {\bibinfo {author} {\bibfnamefont {T.}~\bibnamefont
  {Szidarovszky}}, \bibinfo {author} {\bibfnamefont {A.~G.}\ \bibnamefont
  {Cs\'asz\'ar}}, \bibinfo {author} {\bibfnamefont {G.~J.}\ \bibnamefont
  {Hal\'asz}}, \ and\ \bibinfo {author} {\bibfnamefont {A.}~\bibnamefont
  {Vib\'ok}},\ }\href {\doibase 10.1103/PhysRevA.100.033414} {\bibfield
  {journal} {\bibinfo  {journal} {Phys. Rev. A}\ }\textbf {\bibinfo {volume}
  {100}},\ \bibinfo {pages} {033414} (\bibinfo {year} {2019})}\BibitemShut
  {NoStop}%
\end{thebibliography}%

\end{document}